\documentclass[sigconf]{acmart}
\AtBeginDocument{%
  \providecommand\BibTeX{{%
    \normalfont B\kern-0.5em{\scshape i\kern-0.25em b}\kern-0.8em\TeX}}}

\usepackage{float}

\usepackage{booktabs}
\usepackage{array}
\usepackage{balance} 
\usepackage{multirow}
\usepackage[normalem]{ulem}
\usepackage{color}
\definecolor{lightgray}{RGB}{215,215,215}
\definecolor{myred}{RGB}{210,109,91}
\usepackage{colortbl}  
\usepackage{xcolor}
\useunder{\uline}{\ul}{}
\usepackage{subfigure}
\usepackage{algorithm}  
\usepackage{algorithmicx}  
\usepackage[noend]{algpseudocode}  

\usepackage{amsmath}  
\usepackage{enumitem}
\usepackage{tabularx}
\usepackage[utf8]{inputenc}
\usepackage[english]{babel}
\usepackage{amsthm}
\usepackage{bm}
\newcommand{\ie}{\emph{i.e., }}
\newcommand{\eg}{\emph{e.g., }}

\newcommand{\wrt}{\emph{w.r.t. }}
\newcommand{\cf}{\emph{cf. }}

\newlength\myindent
\floatname{algorithm}{Algorithm}

\copyrightyear{2026}
\acmYear{2026}
\setcopyright{cc}
\setcctype{by}
\acmConference[WWW '26]{Proceedings of the ACM Web Conference 2026}{April 13--17, 2026}{Dubai, United Arab Emirates}
\acmBooktitle{Proceedings of the ACM Web Conference 2026 (WWW '26), April 13--17, 2026, Dubai, United Arab Emirates}
\acmPrice{}
\acmDOI{10.1145/3774904.3792980}
\acmISBN{979-8-4007-2307-0/2026/04}

\begin{document}

\title{Bringing Reasoning to Generative Recommendation Through the Lens of Cascaded Ranking}

\author{Xinyu Lin}
\orcid{https://orcid.org/0000-0002-6931-3182}
\affiliation{%
  \institution{National University of Singapore}
  \city{}
  \country{Singapore}
}
\email{xylin1028@gmail.com}

\author{Pengyuan Liu}
\orcid{https://orcid.org/0009-0001-7639-8815}
\affiliation{%
   \institution{University of Science and Technology of China}
   \city{Hefei}
   \country{China}
}
\email{liupy@mail.ustc.edu.cn}

\author{Wenjie Wang}
\authornote{Corresponding author. This research is supported by the National Natural Science Foundation of China (62572451), the National Research Foundation, Singapore under its National Large Language Models Funding Initiative, (AISG Award No: AISG-NMLP-2024-002). Any opinions, findings and conclusions or recommendations expressed in this material are those of the author(s) and do not reflect the views of National Research Foundation, Singapore.}
\orcid{https://orcid.org/0000-0002-5199-1428}
\affiliation{%
   \institution{University of Science and Technology of China}
   \city{Hefei}
   \country{China}
}
\email{wenjiewang96@gmail.com}

\author{Yicheng Hu}
\orcid{https://orcid.org/0009-0004-8491-8181}
\affiliation{%
   \institution{University of Science and Technology of China}
   \city{Hefei}
   \country{China}
}
\email{huyicheng@mail.ustc.edu.cn}

\author{Chen Xu}
\orcid{https://orcid.org/0000-0002-3070-9358}
\affiliation{%
  \institution{\mbox{Renmin University of China}}
  \city{Beijing}
  \country{China}
}
\email{xc\_chen@ruc.edu.cn}

\author{Fuli Feng}
\orcid{https://orcid.org/0000-0002-5828-9842}
\affiliation{%
  \institution{{University of Science and Technology of China}}
  \city{Hefei}
  \country{China}
}
\email{fulifeng93@gmail.com}

\author{Qifan Wang}
\orcid{https://orcid.org/0000-0002-7570-5756}
\affiliation{
\institution{Meta AI}
\city{Menlo Park}
\country{USA}
}
\email{wqfcr@meta.com}

\author{Tat-Seng Chua}
\orcid{https://orcid.org/0000-0001-6097-7807}
\affiliation{
\institution{National University of Singapore}
\city{}
\country{Singapore}
}
\email{dcscts@nus.edu.sg}

\renewcommand{\shortauthors}{Lin et al.} 

\begin{abstract}

Generative Recommendation (GR) has become a promising end-to-end approach with high FLOPS utilization for resource-efficient recommendation. 
Despite the effectiveness, we show that current GR models suffer from a critical \textbf{bias amplification} issue, where token-level bias escalates as token generation progresses, ultimately limiting the recommendation diversity and hurting the user experience.  
By comparing against the key factor behind the success of traditional multi-stage pipelines, 
we reveal two limitations in GR that can amplify the bias: homogeneous reliance on the encoded history, and fixed computational budgets that prevent deeper user preference understanding. 

To combat the bias amplification issue, it is crucial for GR to 1) 
incorporate more heterogeneous information, and 2) allocate greater computational resources at each token generation step. 
To this end, we propose CARE, 
a simple yet effective cascaded reasoning framework for debiased GR. 
To incorporate heterogeneous information, we introduce a progressive history encoding mechanism, which progressively incorporates increasingly fine-grained history information as the generation process advances. 
To allocate more computations, we propose a query-anchored reasoning mechanism, which seeks to perform a deeper understanding of historical information through parallel reasoning steps. 
We instantiate CARE on three GR backbones. Empirical results on four datasets show the superiority of CARE in recommendation accuracy, diversity, efficiency, and promising scalability. 
The codes and datasets are available at~\url{https://github.com/Linxyhaha/CARE}. 

\end{abstract}


\maketitle

\section{Introduction}\label{sec:intro}

\begin{figure}[t]
\setlength{\abovecaptionskip}{0.02cm}
\setlength{\belowcaptionskip}{-0.3cm}
\centering
\includegraphics[width=0.99\linewidth]{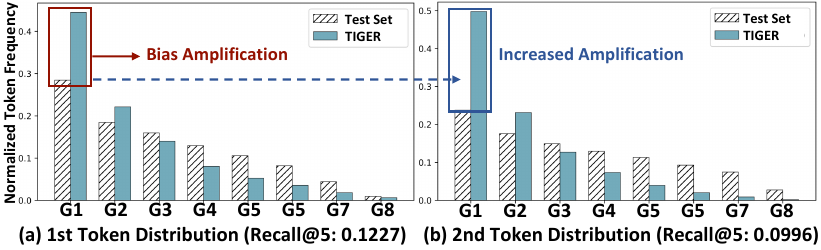}
\caption{Illustration of (1) bias amplification on the popular token groups; and (2) increased bias amplification from initial to later token generation steps (\ie 1st to 2nd token). G1-G8 denotes token groups sorted by popularity. }
\label{fig:bias}
\end{figure}

\begin{figure}[t]
\vspace{0.1cm}
\setlength{\abovecaptionskip}{0.05cm}
\setlength{\belowcaptionskip}{-0.3cm}
\centering
\includegraphics[width=0.99\linewidth]{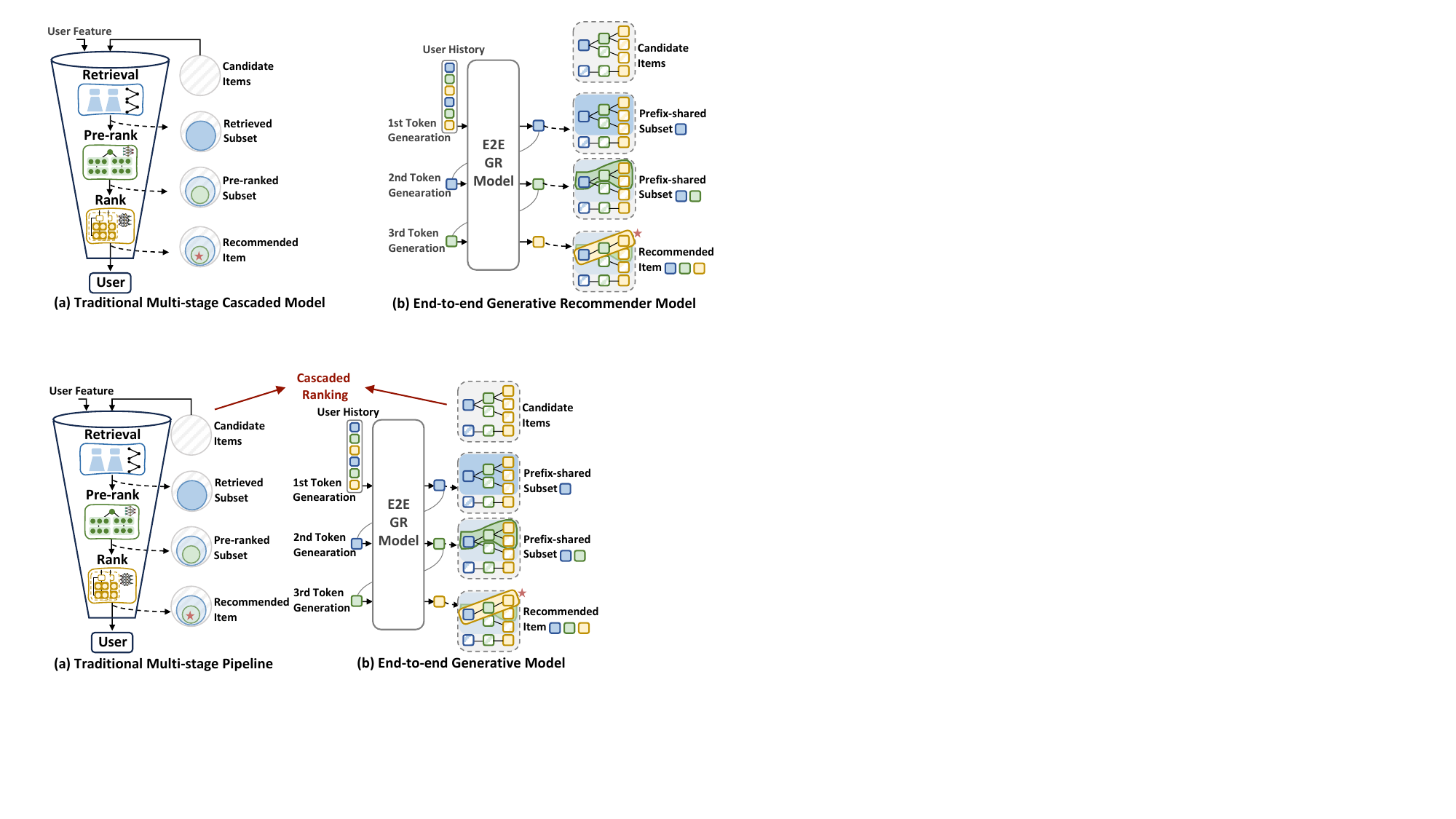}
\caption{Paradigm comparisons between traditional recommendation and generative recommendation, where both essentially perform cascaded item ranking.}
\label{fig:paradigm_comparison}
\end{figure}

Recommender systems have achieved great success in mitigating information overload on the web. 
Nonetheless, the traditional multi-stage pipelines incur substantial communication and storage overhead and suffer from inconsistent optimization objectives across stages~\cite{zhou2025onerec,zhu2025rankmixer}. 
These issues lead to suboptimal recommendation quality and extremely low FLOPs utilization (\eg only 4.6\% in Kuaishou~\cite{zhou2025onerec} and 4.5\% in Douyin~\cite{zhu2025rankmixer}), 
ultimately hindering the development of resource-efficient recommender eco-systems. 
Recently, the development of highly efficient, end-to-end generative recommendation (GR) has emerged as a particularly promising avenue in the recommendation community~\cite{zhou2025onerec,zhu2025rankmixer,he2025plum,rajput2023recommender}.

Technically, GR encodes the user history and directly generates the next item's semantic ID, \ie a token sequence that encodes coarse-grained to fine-grained semantics, to achieve end-to-end recommendation~\cite{rajput2023recommender,he2025plum}. 
Despite the effectiveness, we identify that GR suffers from the critical \textit{\textbf{bias amplification issue}} (Figure~\ref{fig:bias}). 
Specifically, compared with the test distribution, GR tends to assign disproportionately high generation probabilities to frequent tokens, thereby amplifying the \emph{token-level popularity bias}. 
Worse still, such a problem becomes more pronounced as the model progresses from generating the initially coarse-grained to fine-grained semantic tokens (Figure~\ref{fig:bias}(a) to (b)). 
As a result, this amplification of popularity bias continuously accumulates during item token generation and ultimately limits the diversity of recommendations, thereby
degrading the user experience.

To analyze the underlying cause of the issue, we draw inspiration from the multi-stage pipeline with cascaded ranking. 
The key factor behind the success of the multi-stage pipeline is that \textit{pursuing deeper feature interactions to capture more complex user preferences} (Figure~\ref{fig:paradigm_comparison}(a)). 
Its early stage often employs simple and efficient models to capture broad user interests (\eg ANN~\cite{johnson2019billion} for retrieval), while the later stage often adopts more complex models with richer features to fully understand finer-grained interests (\eg deep cross network~\cite{wang2017deep} for ranking). 
On the other hand, GR also implicitly achieves such a ``broad-to-fine-grained'' process with a single unified model via token generation (see Figure~\ref{fig:paradigm_comparison}(b) and details in Section~\ref{sec:preliminaries}).

The cascaded models in the multi-stage pipeline inspire two limitations in GR, which might intensify the bias amplification in the item token generation: 
\begin{itemize}[leftmargin=*]
    \item \textit{\textbf{Homogeneous information}}. 
    Unlike multi-stage pipelines that encode features progressively across stages, GR models heavily rely on the same encoded user's history information (\ie full KV Cache of user history) to generate every subsequent token. 
    As such, the model fails to yield deeper history understanding and struggles to distinguish finer-grained user interests, 
    thus amplifying the superficial influence of frequently observed tokens and introducing strong bias. 
    \item \textit{\textbf{Fixed computation}}. 
    While traditional multi-stage pipelines utilize greater computational effort by more complicated models in later stages, GR models utilize fixed computational resources across all token generation steps through a single forward pass. 
    Consequently, this reduces the model’s ability to discriminate among subtle semantic differences, leading to amplified token bias, especially in the later stage (Figure~\ref{fig:bias}). 
\end{itemize}

Based on these insights, it is crucial for GR to decrease token bias by 1) incorporating more heterogeneous information, and 2) allocating greater computational resources at each token generation step. 
To achieve the first objective, 
each generation step should leverage diverse yet increasingly informative history information, which naturally aligns with the progressive nature of token generation. 
Building on this progressively enriched information, it is promising to harness LLM reasoning~\cite{guo2025deepseek} for subsequent token generation in order to infer finer-grained user preferences through deliberate reasoning steps, which increases the computational budget during inference~\cite{muennighoff2025s1}. 
Nonetheless, simply adding autoregressive reasoning steps can greatly increase the time costs and hurt the practicality of real-time recommendation~\cite{leviathan2023fast}. Hence, it is important to enrich computes while maintaining the time efficiency for practical deployments.

\begin{figure}[t]
\vspace{-0.2cm}
\setlength{\abovecaptionskip}{0.02cm}
\setlength{\belowcaptionskip}{-0.3cm}
\centering
\includegraphics[width=0.99\linewidth]{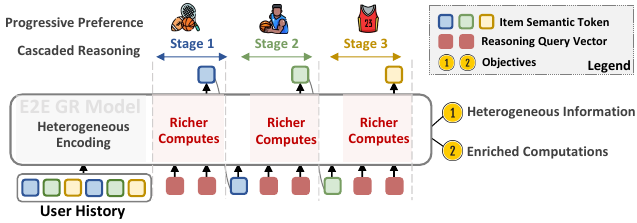}
\caption{Cascaded reasoning framework with two key considerations, \ie heterogeneous information incorporation and enriched computations allocation.}
\label{fig:reasoning_task}
\end{figure}

To this end, we propose \textbf{CARE}, a simple but effective CAscaded REasoning framework for generative recommendation (Figure~\ref{fig:CARE_framework}). 
1) To pursue heterogeneous information for reasoning, 
we devise a progressive history encoding mechanism with a progressive attention mask, which enables the model to depend on richer historical information aligned with semantic granularity as reasoning stages proceed. 
This is motivated by the progressiveness of cascaded ranking, \ie the latter, the more fine-grained information is needed for fine-grained user preference. 
2) To efficiently enrich computations, we introduce a query-anchored reasoning mechanism, which injects multiple model forward passes in parallel before each item token generation. 
To further encourage the model to generate less-frequent tokens, we utilize a diversity loss to regulate the distinctiveness between each reasoning step, aiming to capture diverse user preferences that decrease token generation bias.  
We instantiate CARE on three GR backbones (\ie TIGER, LETTER, and SETRec) on four real-world datasets to demonstrate the effectiveness in ranking accuracy, diversity, efficiency, and reasoning scalability. 

The main contributions of this work are summarized as follows: 
\begin{itemize}[leftmargin=*]
    \item We identify the critical bias amplification issue in GR and examine the causes of homogeneous information and fixed computation. 
    \item We propose a cascaded reasoning framework for GR, named CARE, which aims to leverage more computes for deeper history encoding to infer complex user preferences, achieving diverse and proper recommendations.
    \item Extensive experiments on four real-world datasets demonstrate the accuracy, diversity, efficiency, and promising scalability of our proposed method, unlocking the potential of GR to achieve a more sustainable recommendation ecosystem. 
\end{itemize}
\section{Preliminaries}\label{sec:preliminaries}
In this section, we first introduce generative recommendation and identify the bias amplification issues. We then analyze potential causes of this issue from the perspective of cascaded ranking, and point out the two key considerations to combat the issue. 

\subsection{Retrospect of Generative Recommendation} 
\noindent$\bullet\quad$\textbf{Generative Recommendation}. 
Formally, in GR, each item is represented by a semantic ID denoted as ${\bm{c}}=(c_1, c_2, \dots, c_l)$, \ie a token sequence endowed with hierarchical semantics from coarse- to fine-grained semantics, which is typically obtained by RQ-VAE~\cite{lee2022autoregressive} or hierarchical clustering~\cite{hua2023index}. 
Then, given the user's historical interactions ${X}=[\bm{c}_1, \bm{c}_2, \dots, \bm{c}_L]$, the generative recommender model $\mathcal{M}$ generates the next item's semantic ID, $\bm{c}_{L+1}$, via autoregressive generation: 
\begin{equation}\label{eq:autoregressive_gen}
\small
\left\{
\begin{aligned}   
    &\bm{h}=KV(X) \\
    &c_t = \mathop{\arg\max}_{c}  \mathcal{M}(c|\bm{h},\bm{c}_{<t}), 
\end{aligned}
\right.
\end{equation}
where $\bm{h}$ is the encoded user history, $KV(\cdot)$ is the Key-value computations in standard self-attention mechanism~\cite{vaswani2017attention}, $c_t$ is the $t$-th token of the semantic ID, and $\bm{c}_{<t}$ represents the generated tokens preceding $c_t$. For brevity, we omit the subscript of $L+1$.

\vspace{3pt}
\noindent$\bullet\quad$\textbf{Bias Amplification Issue}. 
Despite the effectiveness of GR in end-to-end item generation, we point out that it suffers from a bias amplification issue. As shown in Figure~\ref{fig:bias}, 1) popularity bias is amplified for each generation step (an average of 246\% times amplification on the most popular token group). When generating each token $c_t$ at time step $t$, generative models will amplify the popularity bias in the semantic tokens, consequently causing item-level bias (\cf Section~\ref{sec:item_generation_distribution}). 
2) Bias amplification becomes more pronounced for the later tokens that capture fine-grained semantics. Comparing the bias amplifications between the coarse-grained first token and the finer-grained second token, the amplification for the popular tokens yields more than 200\%, which further hurts the quality of the generated tokens with almost 20\% accuracy reduction, ultimately limiting the recommendation diversity and exacerbating the echo chamber problem. More empirical evidence is presented in Appendix~\ref{app:bias_amplification}.

\subsection{Diagnose of Bias Amplification Issue}
To examine the underlying causes for the bias amplification issue in GR, we revisit GR through the lens of cascaded ranking, and then identify potential limitations that cause the bias amplification.

\textbf{\textit{Analogy to cascaded ranking}}. 
Although GR generates the next item in an end-to-end manner, it essentially achieves cascaded ranking, where each token generation step in GR is analogous to a specific ranking stage (\eg first token for retrieval and second token for pre-ranking). 
Specifically, at each generation step, the candidate items are the ones that share the same previously generated token sequence (\eg ``<a\_1>''). 
The model then ranks these candidates by scoring possible next tokens (\eg ``<b\_1>'', ``<b\_2>'', ``<b\_3>'') and retains an item subset by selecting the top-scoring token (\eg ``<b\_1>''). Consequently, the candidate set is refined to items consistent with the newly extended prefix (\eg ``<a\_1><b\_1>''), which is passed to the subsequent generation step.

\textbf{\textit{Inherent Limitations}}. 
A key factor behind the success of multi-stage pipelines is to achieve deeper feature interactions to capture more complex user preferences. 
In contrast, unified generative models fail to achieve this due to two limitations. 
1) Homogeneous information. As shown in Eq.(\ref{eq:autoregressive_gen}), the generative model heavily relies on the same history information $X$ for each subsequent token generation, where the model struggles to infer underlying user preference and thus is susceptible to the popularity bias. 
2) Fixed computation. The generative model generates each token by a fixed single forward step (Eq.(\ref{eq:autoregressive_gen})), which limits deeper history interactions and leads to amplified token bias.

\vspace{2pt}
\textit{\textbf{Summary}}. Based on the above insights, it is crucial for GR to reduce the bias via 
1) heterogeneous information incorporation, which emphasizes the removal of dependency on the same encoded history. 
To achieve this, the key lies in the distinctive yet increasingly informative information utilization for encoding. 
2) Enriched computation allocation, which targets injecting more computations into deep information interactions. 
Motivated by the recent success of LLM reasoning, we consider injecting additional reasoning steps for deep user understanding. 

\section{Our Framework: CARE}\label{sec:method}

\begin{figure}[t]
\vspace{-0.3cm}
\setlength{\abovecaptionskip}{0.02cm}
\setlength{\belowcaptionskip}{-0.0cm}
\centering
\includegraphics[width=0.99\linewidth]{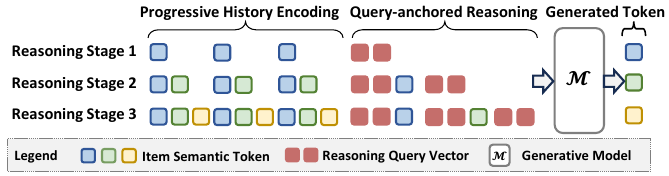}
\caption{Illustration of CARE, including progressive history encoding and query-anchored reasoning.} 
\label{fig:CARE_framework}
\end{figure}

To mitigate the bias amplification issue, we propose a simple yet effective framework CARE, as shown in  Figure~\ref{fig:CARE_framework}. It includes two key components, \ie query-anchored reasoning and progressive history encoding. 
In this section, we first introduce the framework with two components (Section~\ref{sec:framework}), followed by the training of CARE (Section~\ref{sec:training}) and instantiation (Section~\ref{sec:instantiation}).

\subsection{Cascaded Reasoning Architecture}\label{sec:framework}

CARE introduces 1) query-anchored reasoning, and 2) progressive history encoding, to pursue enriched computation allocation and heterogeneous information incorporation, respectively. 

\subsubsection{\textbf{Query-anchored Reasoning}} 
To enable deeper interaction with user history, we borrow the idea of test-time compute in LLMs~\cite{snell2025scaling}, which incorporates additional token generation steps before yielding the final answer and has shown promising results. 
However, directly introducing additional token generation steps requires multiple model forward passes, which can substantially increase inference latency.

In order to retain the benefits of deeper computation without incurring prohibitive runtime costs, we propose query-anchored reasoning that enriches model computations within a single forward pass, denoted as a reasoning stage (Figure~\ref{fig:CARE_framework}). 
Specifically, at each reasoning stage $t$, we introduce a set of learnable query vectors $\bm{Q}_t=(\bm{q_1^t}, \bm{q}_2^t,\dots, \bm{q}_n^t)$, to act as anchors for the deeper preference understanding, where $n$ is the number of query vectors. 
These query vectors are then concatenated to the model input for the token generation. Formally, 
\begin{equation}\label{eq:query_anchored_reasoning}
\small
\left\{
\begin{aligned}
& \bm{p} = \mathcal{M}(\bm{h},\bm{Q}_1,c_{1}, \dots, \bm{Q}_{t-1}, {c}_{t-1}, \bm{Q}_t)\\
    & c_t=\mathop{\arg\max}_{c}\space \bm{p}[c]
\end{aligned}
\right.
\end{equation}
where $\bm{p}\in\mathbb{R}^{|\{c\}|}$ is the generated item token probability corresponding to the last query $\bm{q}_n^t$, $\bm{h}$ is the encoded history, and ${c}_{i\in\{1,\dots,t-1\}}$ is the previously generated item tokens. 
We use the last query to generate the next item token because it effectively integrates all previous reasoning from anchored queries $\bm{q}^t_{1:n-1}$ via the self-attention mechanism.

Despite that the query-anchored reasoning allows additional interactions with user historical interactions, 
it still heavily relies on the pre-encoded user history $\bm{h}$, as shown in Eq(\ref{eq:query_anchored_reasoning}), suffering from the homogeneous information issue. 
To mitigate this, we introduce a progressive history encoding mechanism, as detailed in the following section. 

\subsubsection{\textbf{Progressive History Encoding}} 
To alleviate the homogeneous history encoding issue, we take inspiration from multi-stage pipelines: progressively inject more heterogeneous history encoding as the reasoning stage proceeds. 
Specifically, we aim to gradually incorporate more historical information for the reasoning query vectors to interact with, as the reasoning stage proceeds. 
Then, an important question is what initial information should be used and what additional information should be included as token generation proceeds. 

\vspace{2pt}
\noindent$\bullet\quad$\textbf{Progressive Information Selection}. 
To design a strategy for progressive usage of user's historical interactions, we are motivated to leverage the hierarchical information in semantic IDs, which naturally splits the information into a token sequence with progressive granularity (\eg ``<a\_1>'' reflects the coarse-grained information and becomes more fine-grained with ``<b\_1>'' in a semantic ID ``<a\_1><b\_1><c\_1>''). 

Precisely, we use partial information, denoted as $\tilde{X}$, that aligns with the granularity of the token generation step. 
As shown in Figure~\ref{fig:CARE_framework}, 
in early reasoning stages, we only utilize initial tokens of each historical item to reason the user's broad interests (\eg ``sports products''). 
In later reasoning stages, we progressively include more tokens with finer-grained information to reason the user's deeper preference (\eg ``sports products'' $\rightarrow$ ``basketball shoes'').

\vspace{2pt}
\noindent$\bullet\quad$\textbf{Progressive Encoding}. 
Based on the selected progressive information, the generative model encodes it for reasoning query vectors to attend at each generation step, respectively. 
Formally, at each generation step $t$, we have $\bm{h}_t = KV(\tilde{X}_t)$, 
where $\tilde{X}_t$ is the selected partial history information. The progressively encoded history $\bm{h}_t$ will then be used for queries to interact via Eq.(\ref{eq:query_anchored_reasoning}).

\begin{figure}[t]
\vspace{-0.2cm}
\setlength{\abovecaptionskip}{0.02cm}
\setlength{\belowcaptionskip}{-0cm}
\centering
\includegraphics[width=0.75\linewidth]{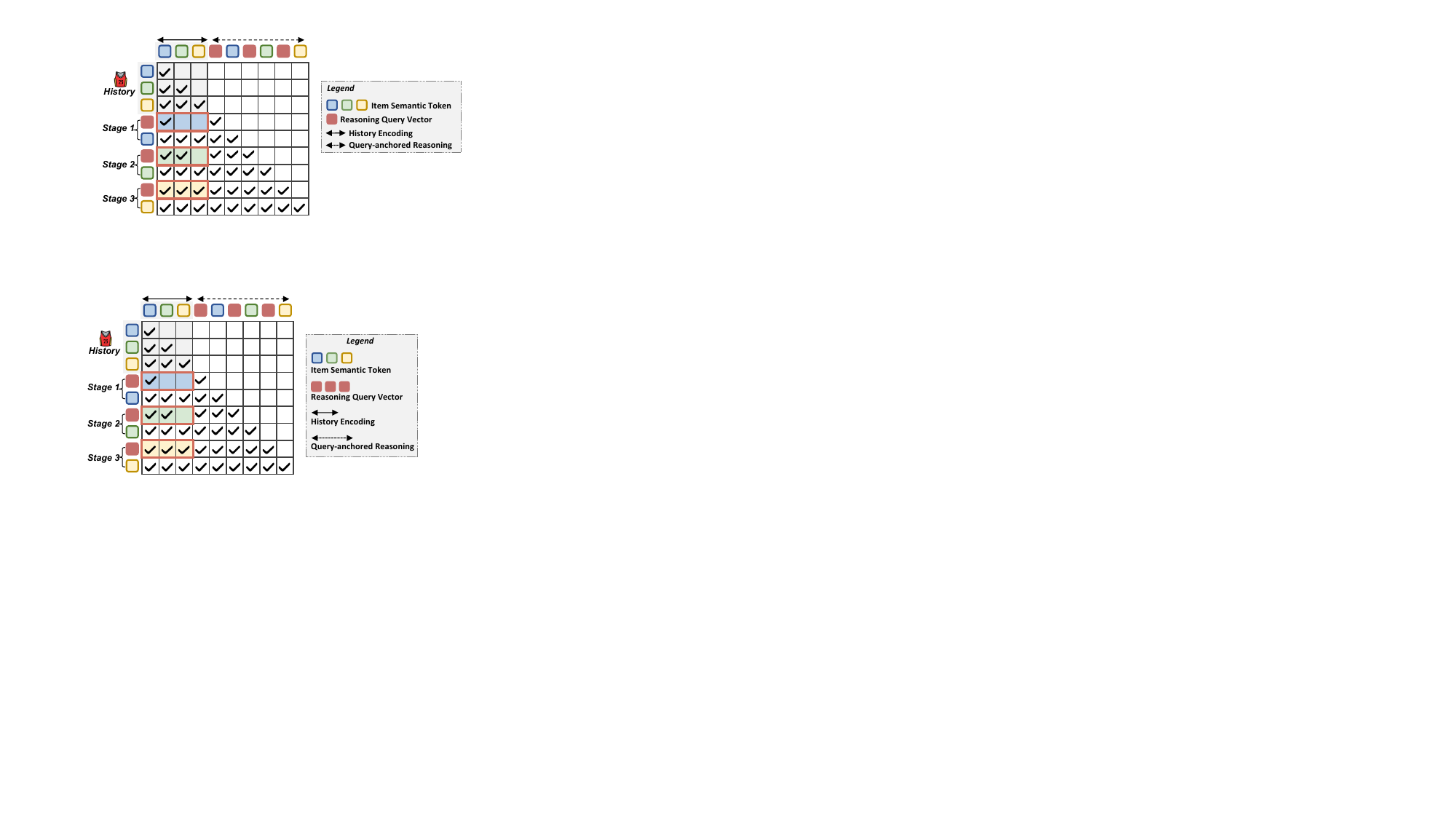}
\caption{Illustration of progressive attention mask, where the query vectors only interact with selected tokens in user history. The example shows one historical item and one query vector for each step. } 
\label{fig:progressive_attention_mask}
\end{figure}

\subsubsection{\textbf{Progressive Attention Mask}}\label{sec:progressive_attention_mask} 
To implement progressive history encoding with standard causal attention mask~\cite{radford2018improving}, an intuitive approach is to encode selective information for each reasoning stage respectively, as shown in Figure~\ref{fig:CARE_framework}. 
Nonetheless, this can cause heavy computation burdens due to the repeated encoding of the history tokens, \eg key-values for the initial tokens will be repeatedly computed across multiple stages. 
To improve computation efficiency, we propose a progressive attention mask. 
It allows the full history to be encoded only once and to be partially interacted with across generation steps, as shown in Figure~\ref{fig:progressive_attention_mask}.

\vspace{2pt}
\noindent$\bullet\quad$\textbf{Time Complexity Analysis on History Encoding}. 
We compare the time complexity of progressive history encoding between our proposed progressive attention mask and the standard causal attention mask. 
With $M$ historical interacted items (each has a semantic ID of length $l$), the standard causal attention mask requires $l$ times history encoding. 
The time complexity for this is then obtained as $\mathcal{O}(M^2d+(2M)^2d+\dots+(lM)^2d) =\mathcal{O}(l^3M^2d)$, where $d$ is the hidden dimension of the backbone LLM.  
In contrast, harnessing the progressive attention mask, we can encode the full history once for all subsequent reasoning stages, which has a time complexity of $\mathcal{O}((lM)^2d)=\mathcal{O}(l^2M^2d)$. 
The time efficiency is improved by reducing repeated computations on the same partial tokens.

\begin{figure}[t]
\vspace{-0.2cm}
\setlength{\abovecaptionskip}{0.0cm}
\setlength{\belowcaptionskip}{-0.3cm}
\centering
\includegraphics[width=0.95\linewidth]{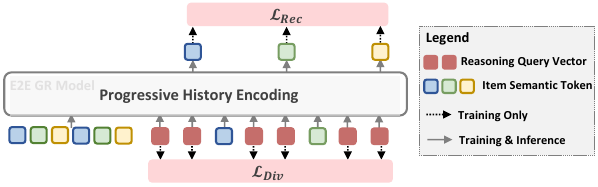}
\caption{Illustration of CARE training via generative recommendation loss and reasoning diversity loss.} 
\label{fig:CARE_optimization}
\end{figure}

\subsection{CARE Training}\label{sec:training} 
To train the generative recommender for deeper user preference understanding, we utilize two loss functions, \ie recommendation loss and reasoning diversity loss, as shown in Figure~\ref{fig:CARE_optimization}.

\vspace{3pt}
\noindent$\bullet\quad$\textbf{Recommendation Loss}. 
To optimize the recommendation ability of the generative model, we adopt the commonly used log-likelihood maximization, defined as: 
\begin{equation}\label{eq:generative_loss}
\small
    \mathcal{L}_{\text{Rec}} = -\sum_{i=1}^{l}\log
    {P}(c_i|\bm{h},\bm{Q}_1,c_{1}, \dots, \bm{Q}_{t-1}, {c}_{t-1}, \bm{Q}_t), 
\end{equation}
where $c_{i\in\{1,\dots,l\}}$ is the token of the next item's semantic ID, and $\bm{Q}_i$ is the query vector set at generation step $i$.

\vspace{3pt}
\noindent$\bullet\quad$\textbf{Reasoning Diversity Loss.} 
In addition to the recommendation loss, we also emphasize the necessity of supervising the reasoning process. Specifically, at each reasoning stage, the reasoning vectors $\bm{Q}$ interact with the same selective attention $\bm{h}_t$, potentially introducing the homogeneous reasoning issue~\cite{zhang2025reinforced} and exacerbating the token bias amplification problem.  
Therefore, we introduce a diversity loss to differentiate the anchoring semantics between queries. Formally, we have:
\begin{equation}\label{eq:diversity_loss}
\small
    \mathcal{L}_{\text{Div}} = \frac{1}{N^{2}-N} \sum_{i=1}^{N}\sum_{j=1,j\neq i}^{N}\mathcal{S}_{ij},
\end{equation}
where $\mathcal{S}_{ij}$ is the cosine similarity between the $i$-th query and $j$-th query, $N$ is the total number of reasoning query vectors. 
Intuitively, we penalize similar queries to encourage them to anchor more heterogeneous semantics for reasoning comprehensive fine-grained user preferences. 

\vspace{3pt}
\noindent$\bullet\quad$\textbf{Overall Loss}. The overall loss to train CARE is defined as
\begin{equation}\label{eq:overall_loss}
\small
    \mathcal{L} = \mathcal{L}_{\text{gen}} + \alpha \mathcal{L}_{\text{div}},
\end{equation}
where $\alpha$ is a hyper-parameter that controls the strength of $L_\text{Div}$.

\subsection{CARE Instantiation}\label{sec:instantiation} 
We instantiate CARE by introducing reasoning query vectors $\bm{Q}$ for each generation step, respectively. 
Specifically, given an arbitrary generative model $\mathcal{M}$, the recommendation data $\mathcal{D}=\{(X,\bm{c})\}$, and the learnable query vectors $\{\bm{Q}\}$, where $X$ is the user historical interactions and $\bm{c}$ is the next item's semantic ID, we aim to train CARE via overall loss (Eq.(\ref{eq:overall_loss})). 
During inference, given the user history $X$, CARE performs query-anchored reasoning to generate the next item's semantic ID via Eq.(\ref{eq:query_anchored_reasoning}). 
\section{Experiments}\label{sec:exp}
We carry out extensive experiments on four real-world datasets to answer the following research questions:  
\textbf{RQ1:} How does our proposed CARE perform compared to traditional and generative models on different decoding mechanisms (\ie autoregressive and parallel decoding)?
\textbf{RQ2:} How does CARE perform on ranking diversity and accuracy at both the token level and item level? 
\textbf{RQ3:} How is the efficiency of CARE, and how do different components of CARE (\ie query-anchored reasoning, progressive attention, diversity loss) affect the performance?

\subsection{Experimental Settings}
\subsubsection{\textbf{Datasets}}
We carry out experiments on four real-world datasets from various domains. Specifically, we use three commonly adopted benchmark datasets from the Amazon Reviews datasets\footnote{\url{https://cseweb.ucsd.edu/~jmcauley/datasets/amazon_v2/}.}
1) \textbf{Games}, 2) \textbf{Sports}, and 3) \textbf{Toys}.
The three Amazon datasets capture extensive user interactions within specific e-commerce product categories. Each product is accompanied by detailed textual metadata, including title, description, category, and brand.
Additionally, we adopt a micro-video recommendation dataset 4) \textbf{MicroLens}\footnote{\url{https://github.com/westlake-repl/MicroLens?tab=readme-ov-file}.} with item titles. 
Following previous work~\cite{bao2024decoding}, we first sort user interactions in temporal order and subsequently split them into training, validation, and test subsets with an 8:1:1 proportion. 

\subsubsection{\textbf{Evaluation Metrics.}}
We employ 1) Recall@$K$ and 2) NDCG@$K$, with $K$ set to 5, 10, and 20. Moreover, we employ 3) DivR@$K$, referred to as diversity ratio~\cite{gao2025sprec}, to measure the percentage of unique items in all recommendations; and 4) ORR@$K$~\cite{gao2025sprec}, which presents the over-recommendation ratio that calculates the percentage of the top-5 recommended items in all recommendations. 

\subsubsection{\textbf{Baselines}}
We compare CARE with traditional and generative models that employ different decoding mechanisms (autoregressive and parallel decoding). Specifically, 
1) \textbf{SASRec}~\cite{kang2018self}, 2) \textbf{GRU4Rec}~\cite{hidasi2016session}, and 3) \textbf{CASER}~\cite{tang2018personalized} are three representative traditional methods for sequential recommendation. 
For autoregressive GR, we compare: 
4) \textbf{TIGER}~\cite{rajput2023recommender} utilizes RQ-VAE and codebooks to map item semantics into discrete token sequences. The resulting identifiers are structured hierarchically, encoding information from coarse-grained to fine-grained levels.
5) \textbf{LETTER}~\cite{wang2024learnable} serves as a SOTA method for item tokenization, which leverages both semantic and CF information during RQ-VAE training to generate multi-dimensional identifiers containing multi-dimensional information 
The parallel GR involves: 
6) \textbf{RPG}~\cite{hou2025generating}  is an efficient generative recommendation model that overcomes the problems of inference latency and limited semantic expression by generating long and unordered semantic IDs in parallel.
7) \textbf{HSTU}~\cite{zhai2024actions} discretizes raw item features into tokens to serve as inputs for generative recommendation. 
8) \textbf{SETRec}~\cite{lin2025order} is one of the SOTA parallel GR approaches, which assigns each item with a set of order-agnostic CF and semantic tokens. 
We also compare with the reasoning method and debiasing methods: 
9) \textbf{ReaRec}~\cite{tang2025think} is a representative reasoning method for sequential recommendation, which employs additional latent reasoning steps before recommendation. We extend this method to GR setting for a fair comparison, denoted as ``ReaRec*''. 
10) \textbf{SPRec}~\cite{gao2025sprec} is a recently proposed debias approach for LLM-based recommendation, which employs SFT and DPO to alleviate the over-recommendation issue.

\subsubsection{\textbf{Implementation Details}}
We employ Qwen2.5-0.5B~\cite{yang2024qwen2} as the LLM backbone. We carefully tune the hyperparameters of the baselines following the suggestions in the original papers. During the training process, depending on the number of learnable parameters, the models are trained for a maximum of $200$ epochs with a batch size of $512$. The learning rate for all models is tuned within $ \{1e^{-3}, 5e^{-4}, 3e^{-4}, 1e^{-5}\}$.
Unless otherwise specified, all experiments are conducted on 8 NVIDIA RTX A100 GPUs. Detailed baseline implementations are provided in Appendix~\ref{app:implementation_detail}.

\subsection{Overall Performance (RQ1)}

In this section, we aim to answer RQ1 by comparing the CARE on both autoregressive GR methods  (Section~\ref{sec:overall_main1}) and parallel GR methods (Section~\ref{sec:overall_main2}).

\begin{table*}[t]
\setlength{\abovecaptionskip}{0.05cm}
\setlength{\belowcaptionskip}{0.2cm}
\caption{Overall performance of baselines and CARE instantiated on autoregressive GR. The bold results highlight the better performance in the comparison between the backbone models with and without CARE. R@K and N@K represent the Recall@K and NDCG@K, respectively.} 
\setlength{\tabcolsep}{2.4mm}{
\resizebox{\textwidth}{!}{
\begin{tabular}{c|l|cccccc|c|l|cccccc}
\toprule
\textbf{Dataset} & \multicolumn{1}{l|}{\textbf{Method}} & \textbf{R@5} & \textbf{R@10} & \multicolumn{1}{l}{\textbf{R@20}} & \textbf{N@5} & \multicolumn{1}{l}{\textbf{N@10}} & \textbf{N@20} & \textbf{Dataset} & \multicolumn{1}{c|}{\textbf{Method}} & \textbf{R@5} & \textbf{R@10} & \textbf{R@20} & \textbf{N@5} & \textbf{N@10} & \multicolumn{1}{c}{\textbf{N@20}} \\ \midrule\midrule
 & SASRec & 0.0316 & 0.0487 & 0.0721 & 0.0221 & 0.0276 & 0.0335 &  & SASRec & 0.0737 & 0.0819 & 0.0927 & 0.0681 & 0.0707 & 0.0734 \\
 & GRU4Rec & 0.0238 & 0.0407 & 0.0658 & 0.0154 & 0.0208 & 0.0271 &  & GRU4Rec & 0.0689 & 0.0793 & 0.0935 & 0.0586 & 0.0619 & 0.0654 \\
 & CASER & 0.0261 & 0.0427 & 0.0672 & 0.0176 & 0.0229 & 0.0291 &  & CASER & 0.0504 & 0.0649 & 0.0811 & 0.0342 & 0.0389 & 0.0429 \\  
 & SPRec & 0.0170 & 0.0244 & 0.0358 & 0.0121 & 0.0145 & 0.0174 &  & SPRec & 0.0353 & 0.0481 & 0.0569 & 0.0174 & 0.0216 & 0.0238 \\ \cline{2-8} \cline{10-16}
 & TIGER & 0.0298 & 0.0503 & 0.0824 & 0.0207 & 0.0273 & 0.0353 &  & TIGER & 0.0803 & 0.0929 & 0.1182 & 0.0704 & 0.0744 & 0.0808 \\
 & \cellcolor[HTML]{EFEFEF}\textcolor{gray!16}{66}\textbf{$+$CARE} & \cellcolor[HTML]{EFEFEF}\textbf{0.0338} & \cellcolor[HTML]{EFEFEF}\textbf{0.0544} & \cellcolor[HTML]{EFEFEF}\textbf{0.0874} & \cellcolor[HTML]{EFEFEF}\textbf{0.0232} & \cellcolor[HTML]{EFEFEF}\textbf{0.0299} & \cellcolor[HTML]{EFEFEF}\textbf{0.0382} &  & \cellcolor[HTML]{EFEFEF}\textcolor{gray!16}{66}\textbf{$+$CARE} & \cellcolor[HTML]{EFEFEF}\textbf{0.0848} & \cellcolor[HTML]{EFEFEF}\textbf{0.0981} & \cellcolor[HTML]{EFEFEF}\textbf{0.1195} & \cellcolor[HTML]{EFEFEF}\textbf{0.0755} & \cellcolor[HTML]{EFEFEF}\textbf{0.0798} & \cellcolor[HTML]{EFEFEF}\textbf{0.0851} \\ \cline{2-8} \cline{10-16} 
 & LETTER & 0.0287 & 0.0503 & 0.0840 & 0.0182 & 0.0250 & 0.0335 &  & LETTER & 0.0806 & 0.1004 & 0.1229 & 0.0652 & 0.0716 & 0.0772 \\
\multirow{-7}{*}{\textbf{Games}} & \cellcolor[HTML]{EFEFEF}\textcolor{gray!16}{66}\textbf{$+$CARE} & \cellcolor[HTML]{EFEFEF}\textbf{0.0323} & \cellcolor[HTML]{EFEFEF}\textbf{0.0532} & \cellcolor[HTML]{EFEFEF}\textbf{0.0897} & \cellcolor[HTML]{EFEFEF}\textbf{0.0213} & \cellcolor[HTML]{EFEFEF}\textbf{0.0279} & \cellcolor[HTML]{EFEFEF}\textbf{0.0371} & \multirow{-7}{*}{\textbf{Sports}} & \cellcolor[HTML]{EFEFEF}\textcolor{gray!16}{66}\textbf{$+$CARE} & \cellcolor[HTML]{EFEFEF}\textbf{0.0810} & \cellcolor[HTML]{EFEFEF}\textbf{0.1015} & \cellcolor[HTML]{EFEFEF}\textbf{0.1241} & \cellcolor[HTML]{EFEFEF}\textbf{0.0659} & \cellcolor[HTML]{EFEFEF}\textbf{0.0725} & \cellcolor[HTML]{EFEFEF}\textbf{0.0782} \\ \midrule\midrule
 & SASRec & 0.0470 & 0.0631 & 0.0803 & 0.0329 & 0.0381 & 0.0424 &  & SASRec & 0.0044 & 0.0069 & 0.0117 & 0.0030 & 0.0038 & 0.0050 \\
 & GRU4Rec & 0.0302 & 0.0420 & 0.0577 & 0.0215 & 0.0253 & 0.0292 &  & GRU4Rec & 0.0042 & 0.0061 & 0.0121 & 0.0025 & 0.0031 & 0.0046 \\
 & CASER & 0.0251 & 0.0375 & 0.0560 & 0.0168 & 0.0208 & 0.0255 &  & CASER & 0.0036 & 0.0075 & 0.0130 & 0.0020 & 0.0032 & 0.0046 \\  
  & SPRec & 0.0227 & 0.0359 & 0.0512 & 0.0132 & 0.0175 & 0.0213 &  & SPRec & 0.0010 & 0.0014 & 0.0018 & 0.0009 & 0.0010 & 0.0011 \\ \cline{2-8} \cline{10-16}
 & TIGER & 0.0419 & 0.0610 & 0.0904 & 0.0285 & 0.0346 & 0.0420 &  & TIGER & 0.0130 & 0.0148 & 0.0166 & 0.0086 & 0.0092 & 0.0096 \\
 & \cellcolor[HTML]{EFEFEF}\textcolor{gray!16}{66}\textbf{$+$CARE} & \cellcolor[HTML]{EFEFEF}\textbf{0.0469} & \cellcolor[HTML]{EFEFEF}\textbf{0.0677} & \cellcolor[HTML]{EFEFEF}\textbf{0.0947} & \cellcolor[HTML]{EFEFEF}\textbf{0.0337} & \cellcolor[HTML]{EFEFEF}\textbf{0.0404} & \cellcolor[HTML]{EFEFEF}\textbf{0.0472} &  & \cellcolor[HTML]{EFEFEF}\textcolor{gray!16}{66}\textbf{$+$CARE} & \cellcolor[HTML]{EFEFEF}\textbf{0.0141} & \cellcolor[HTML]{EFEFEF}\textbf{0.0162} & \cellcolor[HTML]{EFEFEF}\textbf{0.0178} & \cellcolor[HTML]{EFEFEF}\textbf{0.0132} & \cellcolor[HTML]{EFEFEF}\textbf{0.0138} & \cellcolor[HTML]{EFEFEF}\textbf{0.0142} \\ \cline{2-8} \cline{10-16} 
 & LETTER & \multicolumn{1}{r}{0.0429} & \multicolumn{1}{r}{0.0673} & \multicolumn{1}{r}{0.1059} & 0.0300 & \multicolumn{1}{r}{0.0378} & \multicolumn{1}{r|}{0.0475} &  & LETTER & \multicolumn{1}{r}{0.0127} & \multicolumn{1}{r}{0.0146} & \multicolumn{1}{r}{0.0194} & \multicolumn{1}{r}{0.0092} & \multicolumn{1}{r}{0.0098} & 0.0110 \\
\multirow{-7}{*}{\textbf{Toys}} & \cellcolor[HTML]{EFEFEF}\textcolor{gray!16}{66}\textbf{$+$CARE} & \multicolumn{1}{r}{\cellcolor[HTML]{EFEFEF}\textbf{0.0494}} & \multicolumn{1}{r}{\cellcolor[HTML]{EFEFEF}\textbf{0.0759}} & \multicolumn{1}{r}{\cellcolor[HTML]{EFEFEF}\textbf{0.1117}} & \multicolumn{1}{r}{\cellcolor[HTML]{EFEFEF}\textbf{0.0340}} & \multicolumn{1}{r}{\cellcolor[HTML]{EFEFEF}\textbf{0.0425}} & \multicolumn{1}{r|}{\cellcolor[HTML]{EFEFEF}\textbf{0.0515}} & \multirow{-7}{*}{\textbf{MicroLens}} & \cellcolor[HTML]{EFEFEF}\textcolor{gray!16}{66}\textbf{$+$CARE} & \multicolumn{1}{r}{\cellcolor[HTML]{EFEFEF}\textbf{0.0154}} & \multicolumn{1}{r}{\cellcolor[HTML]{EFEFEF}\textbf{0.0167}} & \multicolumn{1}{r}{\cellcolor[HTML]{EFEFEF}\textbf{0.0195}} & \multicolumn{1}{r}{\cellcolor[HTML]{EFEFEF}\textbf{0.0109}} & \multicolumn{1}{r}{\cellcolor[HTML]{EFEFEF}\textbf{0.0113}} & \cellcolor[HTML]{EFEFEF}\textbf{0.0120} \\ \bottomrule
\end{tabular}
}}
\label{tab:overall_performance}
\end{table*}

\subsubsection{\textbf{Performance on Autoregressive GR}}\label{sec:overall_main1}

First, as presented in Table~\ref{tab:overall_performance}, we implement CARE on two representative autoregressive GR models (TIGER and LETTER), denoted as “+CARE,” and evaluate them across all four datasets under varying ranking sizes $K$.
From the experimental results, we can observe that:

\begin{itemize}[leftmargin=*] 
    \item Our method (CARE) generally yields better performance compared to the traditional methods (SASRec, GRU4Rec, and CASER). This is because the GR utilizes RQ-VAE-based semantic IDs, representing an item with hierarchical multiple tokens. This yields two main benefits: 1) it captures multi-dimensional information more effectively, and 2) it accounts for coarse-to-fine variations during generation, aligning better with the autoregressive decoding paradigm. 
    Besides, SPRec shows inferior performance, which might possibly due to the heavy dependency on the quality of the item title~\cite{lin2025order}. The lengthy and noisy item titles might negatively affect the debiasing effectiveness. 
    Detailed analysis among baselines are presented in Appendix~\ref{app:overall_performance_analysis}. 

    \item Our method (CARE) achieves better improvements on both TIGER and LETTER across four datasets, validating the effectiveness of our approach. This is because our method 1) injects more computation via reasoning steps, which facilitates the inference of complex user preferences; and 2) applies a progressive attention mask to the history encoding, enabling the model to capture more fine-grained information in later ranking stages for a deeper user understanding. 
\end{itemize}

\begin{table*}[t]
\setlength{\abovecaptionskip}{0.05cm}
\setlength{\belowcaptionskip}{0.2cm}

\caption{Extension of CARE on parallel GR. The bold results highlight the better performance in the comparison between the backbone models with and without CARE.}
\setlength{\tabcolsep}{2.4mm}{
\resizebox{\textwidth}{!}{
\begin{tabular}{c|l|cccccc|c|l|cccccc}
\toprule
\textbf{Dataset} & \multicolumn{1}{l|}{\textbf{Method}} & \textbf{R@5} & \textbf{R@10} & \multicolumn{1}{l}{\textbf{R@20}} & \textbf{N@5} & \multicolumn{1}{l}{\textbf{N@10}} & \textbf{N@20} & \textbf{Dataset} & \multicolumn{1}{c|}{\textbf{Method}} & \textbf{R@5} & \textbf{R@10} & \textbf{R@20} & \textbf{N@5} & \textbf{N@10} & \multicolumn{1}{c}{\textbf{N@20}} \\ \midrule\midrule
 & RPG & 0.0459 & 0.0653 & 0.0918 & 0.0320 & 0.0383 & 0.0450 & & RPG & 0.0887 & 0.1037 & 0.1216 & 0.0555 & 0.0604 & 0.0649 \\
 & HSTU & 0.0415 & 0.0620 & 0.0918 & 0.0287 & 0.0352 & 0.0427 &  & HSTU & 0.0693 & 0.0806 & 0.0903 & 0.0634 & 0.0671 & 0.0695 \\
 & ReaRec* & 0.0484 & 0.0598 & 0.0727 & 0.0334 & 0.0370 & 0.0402 &  & ReaRec* & 0.0577 & 0.0682 & 0.0836 & 0.0416 & 0.0450 & 0.0488 \\ \cline{2-8} \cline{10-16}
 & SETRec & 0.0480 & 0.0742 & 0.1128 & 0.0322 & 0.0406 & 0.0503 &  & SETRec & 0.0902 & 0.1080 & 0.1311 & 0.0732 & 0.0790 & 0.0848 \\ 
 \multirow{-5}{*}{\textbf{Games}} & \cellcolor[HTML]
 {EFEFEF}\textcolor{gray!16}{66}\textbf{$+$CARE} & \cellcolor[HTML]{EFEFEF}\textbf{0.0500} & \cellcolor[HTML]{EFEFEF}\textbf{0.0760} & \cellcolor[HTML]{EFEFEF}\textbf{0.1141} & \cellcolor[HTML]{EFEFEF}\textbf{0.0332} & \cellcolor[HTML]{EFEFEF}\textbf{0.0415} & \cellcolor[HTML]{EFEFEF}\textbf{0.0511} & \multirow{-5}{*}
 {\textbf{Sports}} & \cellcolor[HTML]{EFEFEF}\textcolor{gray!16}{66}\textbf{$+$CARE} & \cellcolor[HTML]{EFEFEF}\textbf{0.0911} & \cellcolor[HTML]{EFEFEF}\textbf{0.1087} & \cellcolor[HTML]{EFEFEF}\textbf{0.1320} & \cellcolor[HTML]{EFEFEF}\textbf{0.0740} & \cellcolor[HTML]{EFEFEF}\textbf{0.0796} & \cellcolor[HTML]{EFEFEF}\textbf{0.0855} \\ \midrule

\end{tabular}
}}
\label{tab:overall_main2}
\end{table*}

\subsubsection{\textbf{Performance on Parallel GR}}\label{sec:overall_main2} 
In addition to the autoregressive GR method, we further extend CARE to parallel GR methods as shown in Table~\ref{tab:overall_main2}.  
We instantiate CARE on the most competitive SETRec and compare it with parallel GR (RPG, HSTU) and reasoning methods (ReaRec$^*$). 
The results on Toys and MicroLens are in Appendix~\ref{app:parallel_GR}. 
From the results, we derive the following observations: 

\begin{itemize}[leftmargin=*]

   \item Parallel GR methods (RPG, HSTU, SETRec) generally achieve better performance than autoregressive generation methods in Table~\ref{tab:overall_performance}. Since parallel approaches generate multiple tokens independently, this alleviates the error accumulation problem associated with autoregressive generation.
   
   \item Furthermore, among these approaches, those utilizing multiple semantic tokens (SETRec, RPG) outperform single-token methods (HSTU, ReaRec*). This is attributed to the capability of multiple tokens to incorporate richer semantic information and better leverage the rich world knowledge of LLMs. In particular, SETRec achieves the best performance within this group. This is because it not only incorporates semantic information but also jointly learns CF information to leverage similar user behaviors. 

   \item Notably, applying our method to SETRec yields consistent improvements on both datasets, validating the effectiveness and strong generalization ability of our approach across generative models with different architectures. However, we observe that compared to autoregressive GR, the improvement brought by CARE to parallel methods is relatively limited. We speculate that a possible reason is that their item tokens do not strictly follow a coarse-to-fine granularity, and thus may not necessarily require progressive information to assist in modeling user preferences. This also indirectly demonstrates the importance of progressiveness within the cascaded reasoning process.

\end{itemize}

\subsection{Debias Analysis (RQ2)}

\begin{table}[t]
\setlength{\abovecaptionskip}{0.05cm}
\setlength{\belowcaptionskip}{0.2cm}
\caption{Performance comparison regarding diversity ratio (DivR@$K$) and over-recommendation ratio (ORR@$K$).}
\setlength{\tabcolsep}{1mm}{
\resizebox{0.48\textwidth}{!}{
\begin{tabular}{lcccccc}
\toprule
\multicolumn{7}{c}{\textbf{Games}} \\ \midrule
\multicolumn{1}{l|}{} & \textbf{DivR@5$\uparrow$} & \textbf{DivR@10$\uparrow$} & \textbf{DivR@20$\uparrow$} & \textbf{ORR@5$\downarrow$} & \textbf{ORR@10$\downarrow$} & \textbf{ORR@20$\downarrow$} \\ \midrule
\multicolumn{1}{l|}{TIGER} & 0.0183 & 0.0132 & 0.0088 & 0.2278 & 0.1592 & 0.1275 \\
\multicolumn{1}{c|}{
\cellcolor[HTML]{EBFFEB}\textcolor[HTML]{EBFFEB}{66}\textbf{$+$CARE}} & \cellcolor[HTML]{EBFFEB}\textbf{0.0260} & \cellcolor[HTML]{EBFFEB}\textbf{0.0174} & \cellcolor[HTML]{EBFFEB}\textbf{0.0109} & \cellcolor[HTML]{EBFFEB}\textbf{0.1516} & \cellcolor[HTML]{EBFFEB}\textbf{0.1140} & \cellcolor[HTML]{EBFFEB}\textbf{0.0860} \\ \hline
\multicolumn{1}{l|}{\textbf{LETTER}} & 0.0202 & 0.0145 & 0.0094 & 0.2341 & 0.1570 & 0.1173 \\
\multicolumn{1}{c|}{
\cellcolor[HTML]{EBFFEB}\textcolor[HTML]{EBFFEB}{66}\textbf{$+$CARE}} & \cellcolor[HTML]{EBFFEB}\textbf{0.0312} & \cellcolor[HTML]{EBFFEB}\textbf{0.0213} & \cellcolor[HTML]{EBFFEB}\textbf{0.0128} & \cellcolor[HTML]{EBFFEB}\textbf{0.1388} & \cellcolor[HTML]{EBFFEB}\textbf{0.0996} & \cellcolor[HTML]{EBFFEB}\textbf{0.0784} \\ \hline
\end{tabular}
}}
\label{tab:diversity_over-recommendation}
\end{table}
\subsubsection{\textbf{Diversity Analysis}}

To analyze the effectiveness of CARE in alleviating the bias amplification issue, we assess the recommendation diversity and over-recommendation ratio. 
The results on the Games dataset are reported in Table~\ref{tab:diversity_over-recommendation}, and we omit other datasets with similar observations. 
From the results, we can observe that: 
1) CARE consistently improves the recommendation diversity (superior performance \wrt DivR@$K$). This validates the effectiveness of the cascaded reasoning for GR, which allows the model to deeply interact with user history to distinguish between fine-grained user preferences.  
Furthermore, 
2) while the top-5 items from TIGER and LETTER exhibit a 10\%-20\% portion among all recommendations, CARE significantly reduces the bias in popularly recommended items (7\%-15\%). 
Further analysis on distributions of tokens and items is presented in Section~\ref{sec:token_generation_distribution} and~\ref{sec:item_generation_distribution}. 

\begin{figure}[t]
\setlength{\abovecaptionskip}{0cm}
\setlength{\belowcaptionskip}{-0cm}
\centering
\includegraphics[width=1\linewidth]{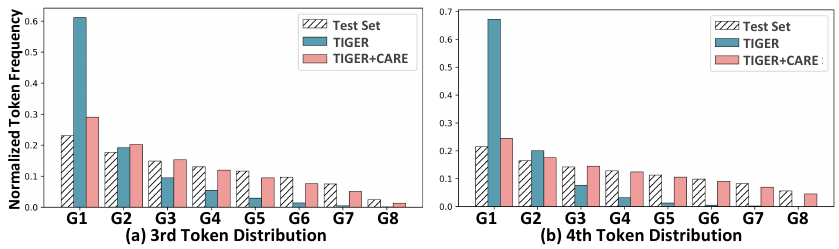}
\caption{Illustration of alleviation of the bias amplification issue by comparing the test distribution and the generated token distributions from TIGER and CARE.}
\label{fig:token_distribution_34}
\end{figure}

\subsubsection{\textbf{Token Generation Distribution}}\label{sec:token_generation_distribution} 
To analyze the effectiveness of our method in mitigating the token-level bias amplification issue, we compare the generated token distributions of the TIGER with and without CARE. Specifically, for each sample, we feed user history together with the next item’s tokens into the generative model and obtain the generated token $c_i$ at each position. 
We then compute the normalized frequency of the generated token $c_i$ at each position and divide them into groups according to the frequency. 
We present the last two token distributions on Games in Figure~\ref{fig:token_distribution_34} and present the other token positions in Appendix~\ref{app:token_distribution}. 

From the experimental results, we observe that:
1) CARE substantially alleviates the token-level bias amplification problem across steps, especially the later tokens, demonstrating the token-level effectiveness of our method. 
2) The mitigation becomes more pronounced, and the generated distribution becomes increasingly close to the ground-truth distribution at later decoding steps. This also indicates that later-stage tokens require deeper and more expressive history encoding to improve ranking diversity.

\subsubsection{\textbf{Item Generation Distribution}}\label{sec:item_generation_distribution} 
\begin{figure}[t]
\setlength{\abovecaptionskip}{0cm}
\setlength{\belowcaptionskip}{-0.3cm}
\centering
\includegraphics[width=0.7\linewidth]{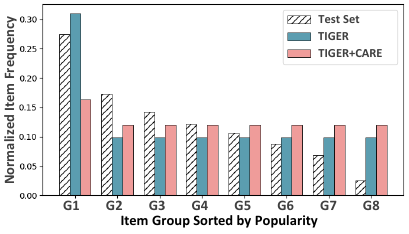}
\caption{Comparison of generated item distributions between TIGER with and without CARE on the Games dataset.}
\label{fig:item_distribution}
\end{figure}

Beyond the token-level analysis, we also examine bias amplification at the item level. We compare the distributions of items generated by TIGER with and without CARE on the Games dataset. As shown in Figure~\ref{fig:item_distribution}, the popular items generated by CARE are significantly less biased compared to both TIGER and the test set. This indicates the potential of CARE in recommending items with better diversity, thereby mitigating the echo chamber issue in recommender systems.

\subsubsection{\textbf{Token Accuracy Analysis}}\label{sec:token_accuracy}

\begin{figure}[t]
\setlength{\abovecaptionskip}{-0.15cm}
\setlength{\belowcaptionskip}{-0cm}
  \centering 
  \hspace{-0.105in}
  \subfigure{
    \includegraphics[width=0.48\linewidth]{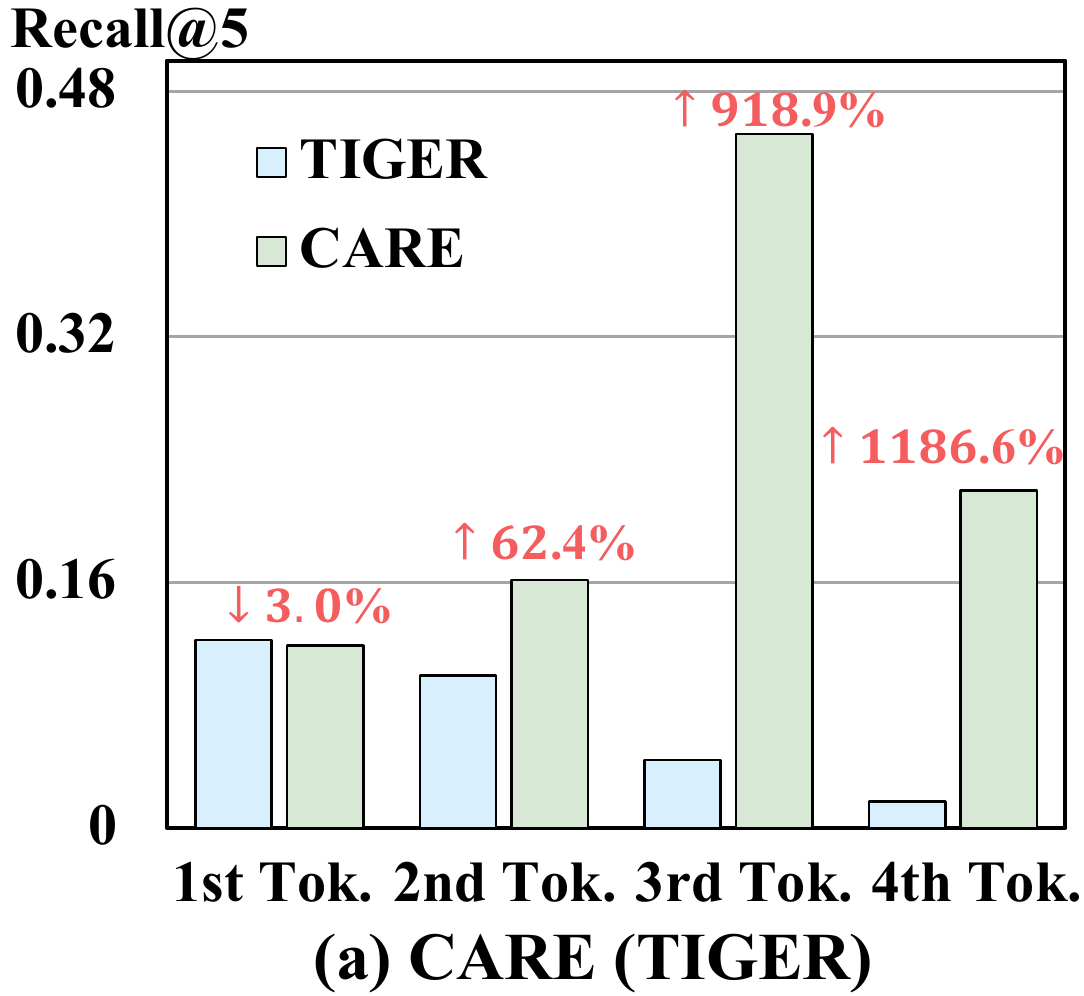}} 
  \hspace{-0.105in}
  \subfigure{
    \includegraphics[width=0.48\linewidth]{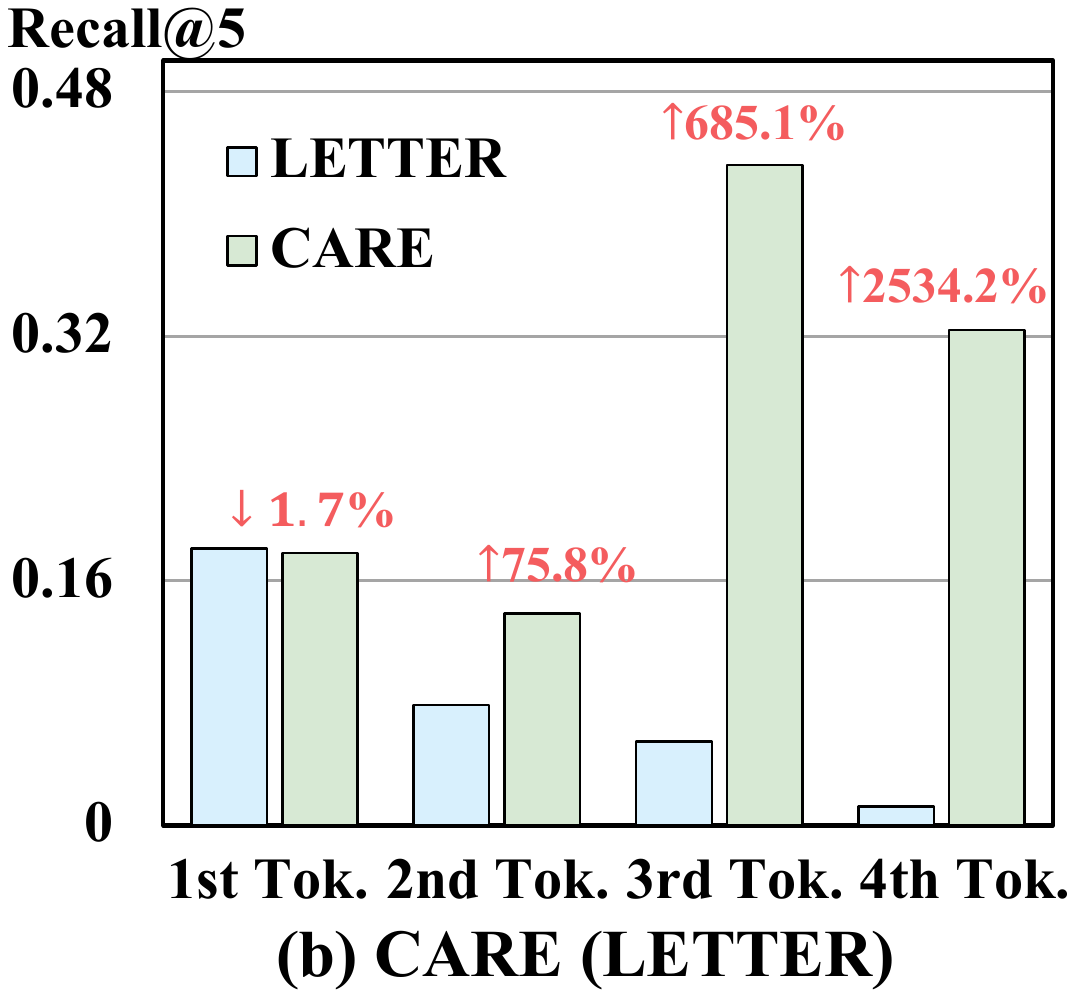}} 
\caption{Recall of each token of CARE on Games dataset.}
  \label{fig:Token_Accuracy}
\end{figure}

To further analyze whether our method can generate more accurate tokens, we follow the same setting in Section~\ref{sec:token_generation_distribution} and calculate the Recall@5 of each token as shown in Figure~\ref{fig:Token_Accuracy}. 
The results show that 
1) CARE significantly improves the generation accuracy on the last three tokens, which partially explains the superiority of CARE in Table~\ref{tab:overall_performance}. 
2) CARE shows comparable Recall on the first token, but the improvement becomes more pronounced with CARE in later generation steps. 
This suggests that in the early stages, historical information might be redundant to capture broad user preferences (refer to Appendix~\ref{app:token_distribution} for further first token analysis).


\subsection{Efficiency and Effectiveness (RQ3)}
We further analyze the efficiency of CARE and the effectiveness of each component, including the reasoning query vector, progressive attention, and diversity loss.

\subsubsection{\textbf{Efficiency Analysis}} 
\begin{table}[t]
\setlength{\abovecaptionskip}{0.05cm}
\setlength{\belowcaptionskip}{0.2cm}
\caption{Time costs and GPU memory consumption between GR backbone and CARE tested on 8 NVIDIA RTX A100 GPUs. ``Tr'' and ``Inf'' denote ``Training'' and ``Inference'', respectively.}
\setlength{\tabcolsep}{2.2mm}{
\resizebox{0.48\textwidth}{!}{
\begin{tabular}{lcccc}
\toprule
 \multicolumn{5}{c}{\textbf{Games}} \\ \midrule
\multicolumn{1}{l|}{} & \textbf{Time (Tr)} & \textbf{Time (Inf)} & \textbf{GPU Mem (Tr)} & \multicolumn{1}{c}{\textbf{GPU Mem (Inf)}} \\ \midrule
\multicolumn{1}{l|}{TIGER} & 21min42s & 639.12s & 23.07GiB/GPU & \multicolumn{1}{c}{9,929MiB} \\
\multicolumn{1}{c|}{\cellcolor[HTML]{ecf4ff}\textcolor[HTML]{ecf4ff}{66}\textbf{$+$CARE}} 
& \cellcolor[HTML]{ecf4ff}22min35s & \cellcolor[HTML]{ecf4ff}652.11s & \cellcolor[HTML]{ecf4ff}24.54GiB/GPU & \multicolumn{1}{c}{\cellcolor[HTML]{ecf4ff}9,963MiB} \\ \hline
\multicolumn{1}{l|}{LETTER} & 22min23s & 523.42s & 23.07GiB/GPU & \multicolumn{1}{c}{9,913MiB} \\
\multicolumn{1}{c|}{\cellcolor[HTML]{ecf4ff}\textcolor[HTML]{ecf4ff}{66}\textbf{$+$CARE}} & \cellcolor[HTML]{ecf4ff}23min42s & \cellcolor[HTML]{ecf4ff}{529.27s} & \cellcolor[HTML]{ecf4ff}24.54GiB/GPU & \multicolumn{1}{c}{\cellcolor[HTML]{ecf4ff}9,947MiB} \\ \hline
\end{tabular}
}}
\label{tab:Efficiency_Analysis}
\end{table}
 
To analyze the efficiency of CARE, we compare the time costs and GPU memory consumption of TIGER and LETTER with and without CARE for both training and inference. 
We set the training batch size to 1024. And the batch size and beam size for inference are set to 50 and 20, respectively. 
The results on the Games dataset are shown in Table~\ref{tab:Efficiency_Analysis}.

We can observe that CARE demonstrates comparable time and memory efficiency.  
For time costs, CARE yields 4.9\% and 1.5\% average increase in training and inference time; For GPU memory usage, CARE shows 6.37\% and 0.34\% average increase in training and inference GPU memory. 
This is attributed to the parallel reasoning that eliminates the need for additional model forward passes. Furthermore, since the user history constitutes the dominant portion of memory consumption, the additional memory usage introduced by the query is minimal. Analysis of the comparison between two backbones is in Appendix~\ref{app:efficiency_analysis}.

\subsubsection{\textbf{Ablation Study}}\label{sec:ablation}

\begin{figure}[t]
\setlength{\abovecaptionskip}{-0.15cm}
\setlength{\belowcaptionskip}{-0cm}
  \centering 
  \hspace{-0.105in}
  \subfigure{
    \includegraphics[width=0.49\linewidth]{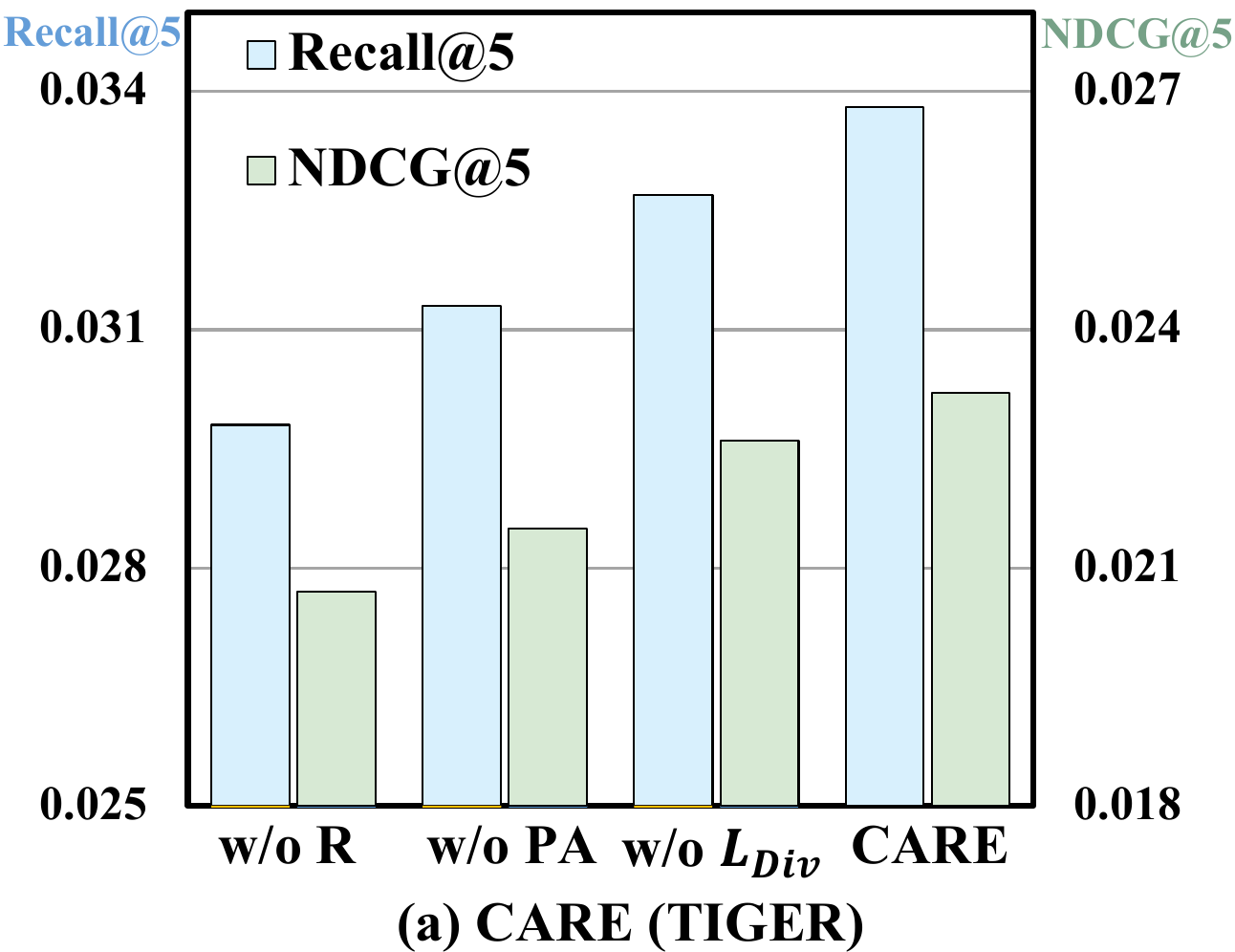}} 
  \subfigure{
    \includegraphics[width=0.49\linewidth]{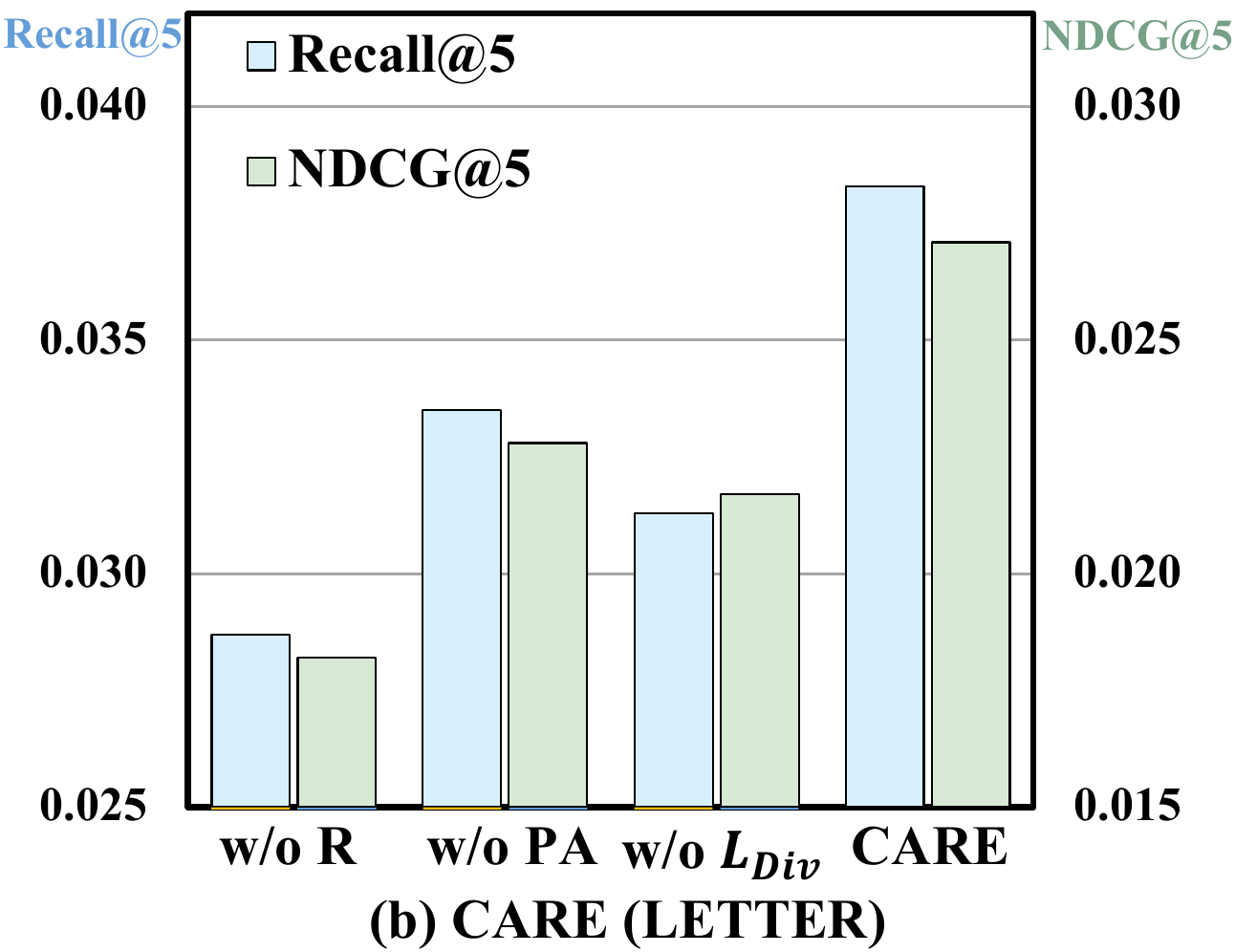}} 
    \caption{Ablation study on Games dataset. ``R'' and ``PA'' denotes ``Reasoning'' and ``Progressive Attention'', respectively.}
  \label{fig:Ablation_analysis1}
\end{figure}

To verify the effectiveness of each component, we individually remove the query-anchored reasoning (``w/o reasoning''), replace the progressive attention mask with the original causal attention mask (``w/o PA''), and remove the diversity loss (``w/o $\mathcal{L}_{\text{Div}}$''). 
We present results on Games in Figure~\ref{fig:Ablation_analysis1} and find that 
1) discarding each component leads to a performance drop, validating their effectiveness. 
2) In particular, the impact of reasoning is the most significant 
because it is the key component to enable deeper interactions with heterogeneous information. 
Further analysis on diversity loss strength is presented in Appendix~\ref{app:effect_of_diversity}.

\subsubsection{\textbf{Scalability on Reasoning Steps}} To investigate the scalability of CARE regarding the number of reasoning steps, we vary the reasoning query number for four generation steps. From the results in Table~\ref{tab:reasoning_scalability}, we can observe that: 
1) Scaling reasoning at a later stage (bottom block in the Table) is generally more effective than early stage (top block). 
This is expected, as the generation of later tokens requires a more refined and fine-grained understanding of user preferences (\cf Section~\ref{sec:preliminaries}).  
2) Injecting more reasoning at the early stage shows limited improvements. The possible reason is that the early stage captures the broad user interests, which is less necessary to require substantial reasoning. This also explains the observation that initial token information is sufficient for generating the coarse-grained item token (\cf Section~\ref{sec:ablation}).

\begin{table}[t]
\setlength{\abovecaptionskip}{0.0cm}
\setlength{\belowcaptionskip}{0.2cm}
\caption{Performance comparison between CARE (TIGER) with different reasoning query numbers on Games dataset.}
\setlength{\tabcolsep}{3.5mm}{
\resizebox{0.48\textwidth}{!}{
\begin{tabular}{l|cccccc}
\toprule
\textbf{\# Query} & \multicolumn{1}{l}{\textbf{R@5}} & \multicolumn{1}{l}{\textbf{R@10}} & \multicolumn{1}{l}{\textbf{R@20}} & \multicolumn{1}{l}{\textbf{N@5}} & \multicolumn{1}{l}{\textbf{N@10}} & \multicolumn{1}{l}{\textbf{N@20}} \\ \midrule
\textbf{2-2-1-1} & 0.0334 & 0.0536 & 0.0907 & 0.0220 & 0.0285 & 0.0378 \\
\textbf{2-4-1-1} & 0.0299 & 0.0515 & 0.0881 & 0.0197 & 0.0266 & 0.0358 \\
\textbf{4-4-1-1} & 0.0320 & 0.0538 & 0.0899 & 0.0218 & 0.0287 & 0.0378 \\
\midrule
\textbf{1-1-2-2} & 0.0313 & 0.0490 & 0.0812 & 0.0221 & 0.0278 & 0.0358 \\
\textbf{1-1-2-4} & 0.0338 & 0.0538 & 0.0910 & 0.0225 & 0.0289 & 0.0382 \\
\textbf{1-1-4-4} & 0.0357 & 0.0555 & 0.0882 & 0.0248 & 0.0312 & 0.0393 \\ \bottomrule
\end{tabular}
}}
\label{tab:reasoning_scalability}
\end{table}

\section{Related Work}\label{sec:related_work}

\noindent$\bullet\quad$\textbf{Generative Recommendation}. 
In recommendation tasks~\cite{qiu-etal-2025-measuring,zhao2025model,igd,liu2025recoworld}, traditional multi-stage pipelines suffer from extremely low model FLOPs utilization (MFU) and the difficulty of optimizing multiple stages jointly~\cite{zhu2025rankmixer,zhou2025onerec,he2025plum}. To better utilize computational resources, GR has emerged as a promising end-to-end paradigm. 
GR typically represents each item with a semantic ID, encoding semantics from coarse-grained to fine-grained levels~\cite{wu2025constrained}. 
Given a user’s historical interaction sequence, the GR model autoregressively generates the next semantic ID tokens.

\vspace{2pt}
\noindent$\bullet\quad$\textbf{Bias Amplification in Generative Recommendation}. 
Popularity bias has been recently explored in natural language-based GR~\cite{dai2024bias, gallegos2024bias, zhang2023chatgpt, zhou2025exploring, sakib2024challenging}, \ie generating the next item's textual identifier, such as title.  
They mainly focus on token-level and item-level bias in natural language~\cite{jiang2024item, bao2024decoding, li2025llm}. 
To address the bias, they typically either rely on prior knowledge of target distributions or external guidance for LLM alignment~\cite{lu2025dual, hou2025leadfairrec}. 
Nonetheless, the bias issue in semantic-ID-based GR receives little scrutiny. 
In this work, we identify the critical token-level bias and take an initial step toward addressing the popularity bias issue through the lens of cascaded ranking. 
While these approaches fail to be directly applied to GR due to the intrinsic gap between the textual tokens and semantic tokens, we discuss and compare a representative work SPRec~\cite{gao2025sprec} under their initial setting (\eg item title generation).

\vspace{2pt}
\noindent$\bullet\quad$\textbf{Reasoning in Recommendation}. 
Test-time Reasoning has shown promising results across various domains~\cite{guo2025deepseek,huang2025vchain}. Existing studies on reasoning for recommendation typically perform a reason-then-recommend paradigm~\cite{bismay2025reasoningrec, liu2025improving, gu2025r, yue2025cot4rec, wang2025re2llm, liu2025lares, zhang2025slow, zhao2025reason,tang2025think}, where additional autoregressive generation steps are performed before generating the final item. 
The incorporation of these deliberate reasoning steps aims to enhance the model’s ability to understand complex user interests, particularly improving performance for long-tail users with sparse interaction histories~\cite{tang2025think}. 
In contrast, our work harnesses reasoning from a novel debiasing perspective. 
Our proposed CARE leverages query-anchored reasoning to achieve deeper history interaction for every token generation, while avoiding the additional time costs raised by autoregressive reasoning. 
To demonstrate our advantage compared with prior reasoning-based approaches, we discuss and compare with the representative reasoning method ReaRec~\cite{tang2025think} in Section~\ref{sec:exp}. 
\section{Conclusion and Future Work}\label{sec:conclusion}
In this work, we revisited GR and identified bias amplification as a core limitation of existing GR models. 
By analyzing the success behind the traditional multi-stage pipeline, we uncovered the two potential causes of the bias amplification issue: homogeneous information utilization and fixed computational budgets across token generation steps. 
To address these limitations, we proposed CARE, a cascaded reasoning framework that introduced progressive history encoding with a progressive attention mask for heterogeneous information and query-anchored reasoning for efficient compute enrichment. 
Experiments on three GR backbones and four datasets verified that CARE improved accuracy, diversity, and efficiency, and showed promising scalability. 

CARE offers new opportunities for diverse recommendation via reasoning, which leaves many promising directions to explore. 
First, dynamic compute allocation may balance reasoning effort across different reasoning stages in large-scale deployments. 
Second, adaptive history selection, especially for the initial token generation, deserves careful design to achieve both accurate and diverse item generation. 
Third, studying the long-term behavioral impact of debiased GR could provide insights into mitigating echo chambers and improving fairness in practical systems. 
\appendix
\section{Appendix}\label{sec:appendix}

\subsection{\textbf{Additional Empirical Evidence of Bias Amplification}}\label{app:bias_amplification} 
\begin{figure}[h]
\vspace{-0.2cm}
\setlength{\abovecaptionskip}{-0.15cm}
\setlength{\belowcaptionskip}{-0cm}
  \centering 
  \hspace{-0.105in}
  \subfigure{
    \includegraphics[width=0.48\linewidth]{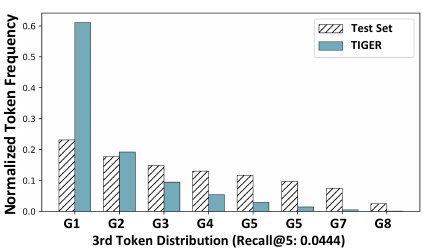}} 
  \hspace{-0.105in}
  \subfigure{
    \includegraphics[width=0.48\linewidth]{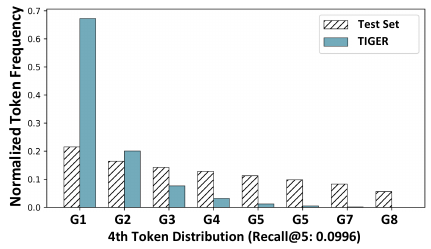}} 
\caption{Illustration of bias amplification on the popular token groups of the 3rd and 4th token on Games dataset.}
  \label{fig:bias_token_3_4}
\end{figure}

Figure~\ref{fig:bias_token_3_4} illustrates the group token distribution of the third and fourth tokens on Games dataset, where we have similar observation as stated in Section~\ref{sec:intro}: bias amplification issue. 
We highlight that the issue becomes more severe in the later tokens. Specifically, the normalized frequency of the most popular token group is amplified by 160\%, 200\%, 300\%, and 325\% times from first to the last tokens, respectively.

\subsection{\textbf{Additional Experimental Results}}

\subsubsection{\textbf{User Group Analysis}}\label{app:app_user_group} 
To further investigate whether our method can better capture complex user preferences, particularly for users with fewer interactions, we group users based on their interaction counts into four categories from Group 1 to Group 4, representing users from the densest to the sparsest interaction histories. On each of these user groups, we then evaluate two representative generative models, TIGER and LETTER, as well as their counterparts equipped with our reasoning framework. The results on the Games dataset are reported in Figure~\ref{fig:user_group}. We omit results from other datasets that exhibit similar trends to save space.

From the figures, we observe that:
1) All methods show a consistent performance decline from Group 1 to Group 4, which is reasonable because it is inherently more challenging to predict user preferences with sparser and more limited interaction history. Also, LETTER generally outperforms TIGER among the baselines, primarily because TIGER’s semantic IDs embed CF signals, relying on behaviors from similar users to infer individual user preferences. A similar trend can also be observed in Table~\ref{tab:overall_performance} of the main results.
2) Our method consistently improves performance across all user groups. This verifies the effectiveness of CARE in enhancing preference modeling under varying levels of user sparsity.

\begin{table*}[h]
\setlength{\abovecaptionskip}{0.05cm}
\setlength{\belowcaptionskip}{0.2cm}
\caption{Performance comparison between baselines and CARE instantiated on parallel GR on Toys and MicroLens datasets. The bold results highlight the better performance in the comparison between the backbone models with and without CARE.} 
\setlength{\tabcolsep}{2.4mm}{
\resizebox{\textwidth}{!}{
\begin{tabular}{c|l|cccccc|c|l|cccccc}
\toprule
\textbf{Dataset} & \multicolumn{1}{l|}{\textbf{Method}} & \textbf{R@5} & \textbf{R@10} & \multicolumn{1}{l}{\textbf{R@20}} & \textbf{N@5} & \multicolumn{1}{l}{\textbf{N@10}} & \textbf{N@20} & \textbf{Dataset} & \multicolumn{1}{c|}{\textbf{Method}} & \textbf{R@5} & \textbf{R@10} & \textbf{R@20} & \textbf{N@5} & \textbf{N@10} & \multicolumn{1}{c}{\textbf{N@20}} \\ \midrule\midrule
  & RPG & 0.065 & 0.0842 & 0.1077 & {0.0493} & {0.0555} & {0.0614} & & RPG & 0.0104 & 0.0133 & 0.0247 & 0.0061 & 0.0069 & 0.0097 \\
  & HSTU & 0.0527 & 0.0662 & 0.0823 & 0.0385 & 0.0428 & 0.0469 &  & HSTU & 0.0098 & 0.0142 & 0.0203 & 0.0063 & 0.0077 & 0.0092 \\
  & ReaRec* & 0.0167 & 0.0263 & 0.0422 & 0.0112 & 0.0143 & 0.0183 &  & ReaRec* & 0.0022 & 0.0027 & 0.0046 & 0.0014 & 0.0015 & 0.0020 \\ \cline{2-8} \cline{10-16}
  & SETRec & \multicolumn{1}{r}{0.0560} & \multicolumn{1}{r}{0.0866} & \multicolumn{1}{r}{0.1234} & 0.0358 & \multicolumn{1}{r}{0.0456} & \multicolumn{1}{r|}{0.0549} &  & SETRec & \multicolumn{1}{r}{0.0121} & \multicolumn{1}{r}{0.0168} & \multicolumn{1}{r}{0.0319} & \multicolumn{1}{r}{0.0068} & \multicolumn{1}{r}{0.0083} & {0.0120} \\
  \multirow{-5}{*}{\textbf{Toys}} & \cellcolor[HTML]{EFEFEF}\textcolor{gray!16}{66}\textbf{$+$CARE} & \multicolumn{1}{r}{\cellcolor[HTML]{EFEFEF}\textbf{0.0585}} & \multicolumn{1}{r}{\cellcolor[HTML]{EFEFEF}\textbf{0.0878}} & \multicolumn{1}{r}{\cellcolor[HTML]{EFEFEF}\textbf{0.1243}} & \multicolumn{1}{r}{\cellcolor[HTML]{EFEFEF}{\textbf{0.0359}}} & \multicolumn{1}{r}{\cellcolor[HTML]{EFEFEF}{0.0453}} & \multicolumn{1}{r|}{\cellcolor[HTML]{EFEFEF}{0.0545}} & \multirow{-5}{*}{\textbf{MicroLens}} & \cellcolor[HTML]{EFEFEF}\textcolor{gray!16}{66}\textbf{$+$CARE} & \multicolumn{1}{r}{\cellcolor[HTML]{EFEFEF}\textbf{0.0126}} & \multicolumn{1}{r}{\cellcolor[HTML]{EFEFEF}\textbf{0.0172}} & \multicolumn{1}{r}{\cellcolor[HTML]{EFEFEF}\textbf{0.0284}} & \multicolumn{1}{r}{\cellcolor[HTML]{EFEFEF}\textbf{0.0069}} & \multicolumn{1}{r}{\cellcolor[HTML]{EFEFEF}\textbf{0.0084}} & \cellcolor[HTML]{EFEFEF}{0.0112} \\ \bottomrule
\end{tabular}
}}
\label{tab:overall_main2_toys_microlens}
\end{table*}

\subsubsection{\textbf{Token Distribution of First and Second Tokens}}\label{app:token_distribution}
\begin{figure}[h]
\setlength{\abovecaptionskip}{0.02cm}
\setlength{\belowcaptionskip}{-0.3cm}
\centering
\includegraphics[width=0.99\linewidth]{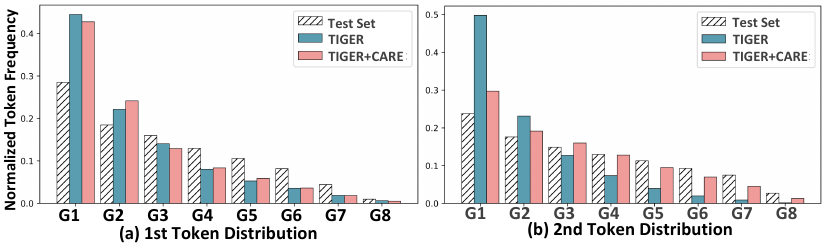}
\caption{Illustration of alleviation of the bias amplification issue by comparing the test distribution, the generated token distributions from TIGER and CARE.}
\label{fig:token_distribution_12}
\end{figure}

In accordance with Section~\ref{sec:token_generation_distribution}, additional token distributions of first and second token are presented in Figure~\ref{fig:token_distribution_12}. We can find that: 
1) CARE alleviates the first and second tokens' amplified bias, which is similar to that of the third and last token (\cf Section~\ref{sec:token_generation_distribution}). 
Nonetheless, 
2) it is noted that the mitigation of bias amplification on later tokens is more effective than that on the first token. 
One possible reason could be temporal user preference shifts, which increase the difficulty of generating the first token accurately, regardless of the information involved in reasoning. 
In light of this, we believe the first token generation requires careful design of reasoning to capture the significant potential user preference shift, which we leave as future work to explore.

\begin{figure}[h]
\vspace{-0.2cm}
\setlength{\abovecaptionskip}{-0.15cm}
\setlength{\belowcaptionskip}{-0cm}
  \centering 
  \hspace{-0.105in}
  \subfigure{
    \includegraphics[width=0.48\linewidth]{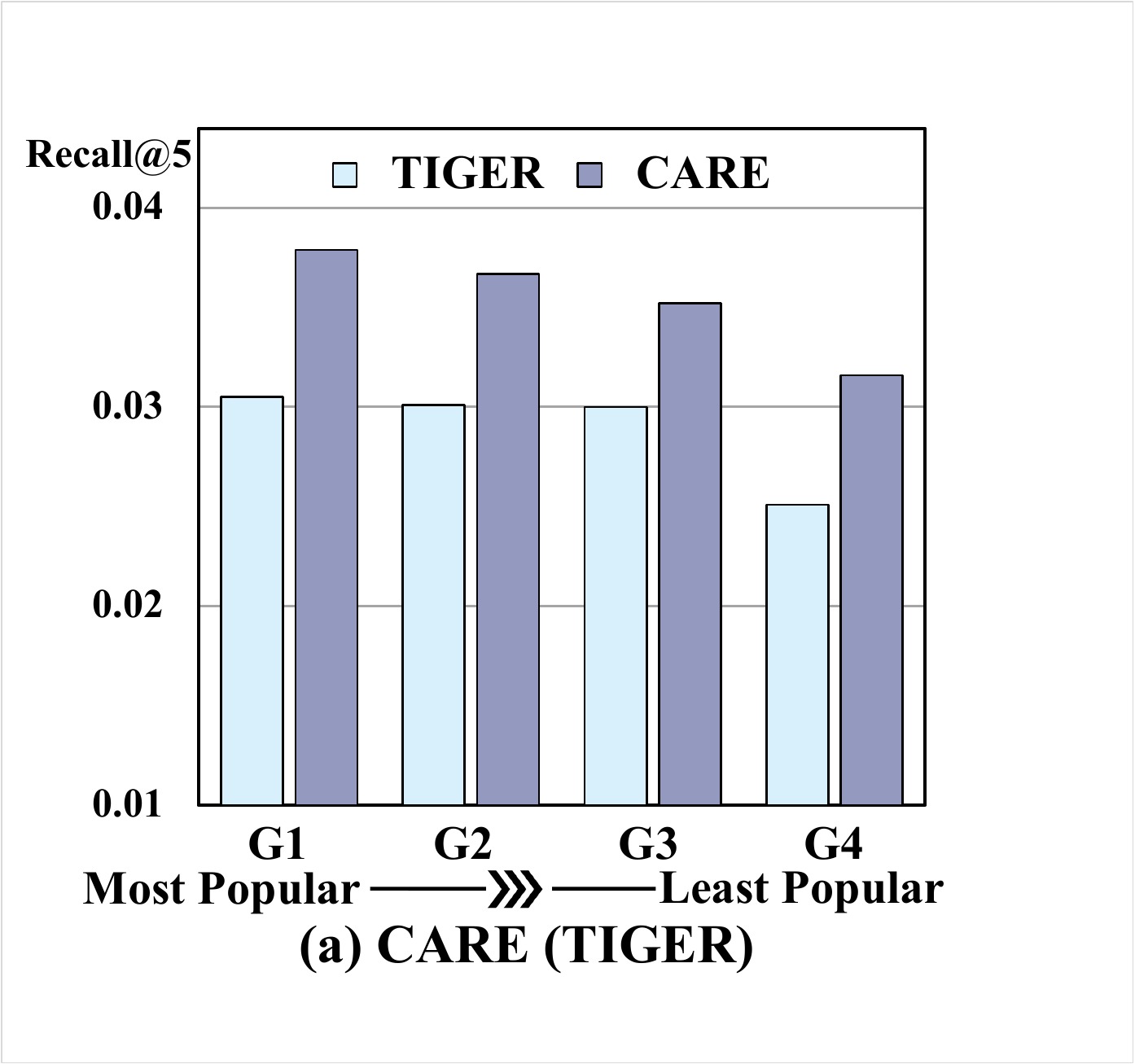}} 
  \hspace{-0.105in}
  \subfigure{
    \includegraphics[width=0.48\linewidth]{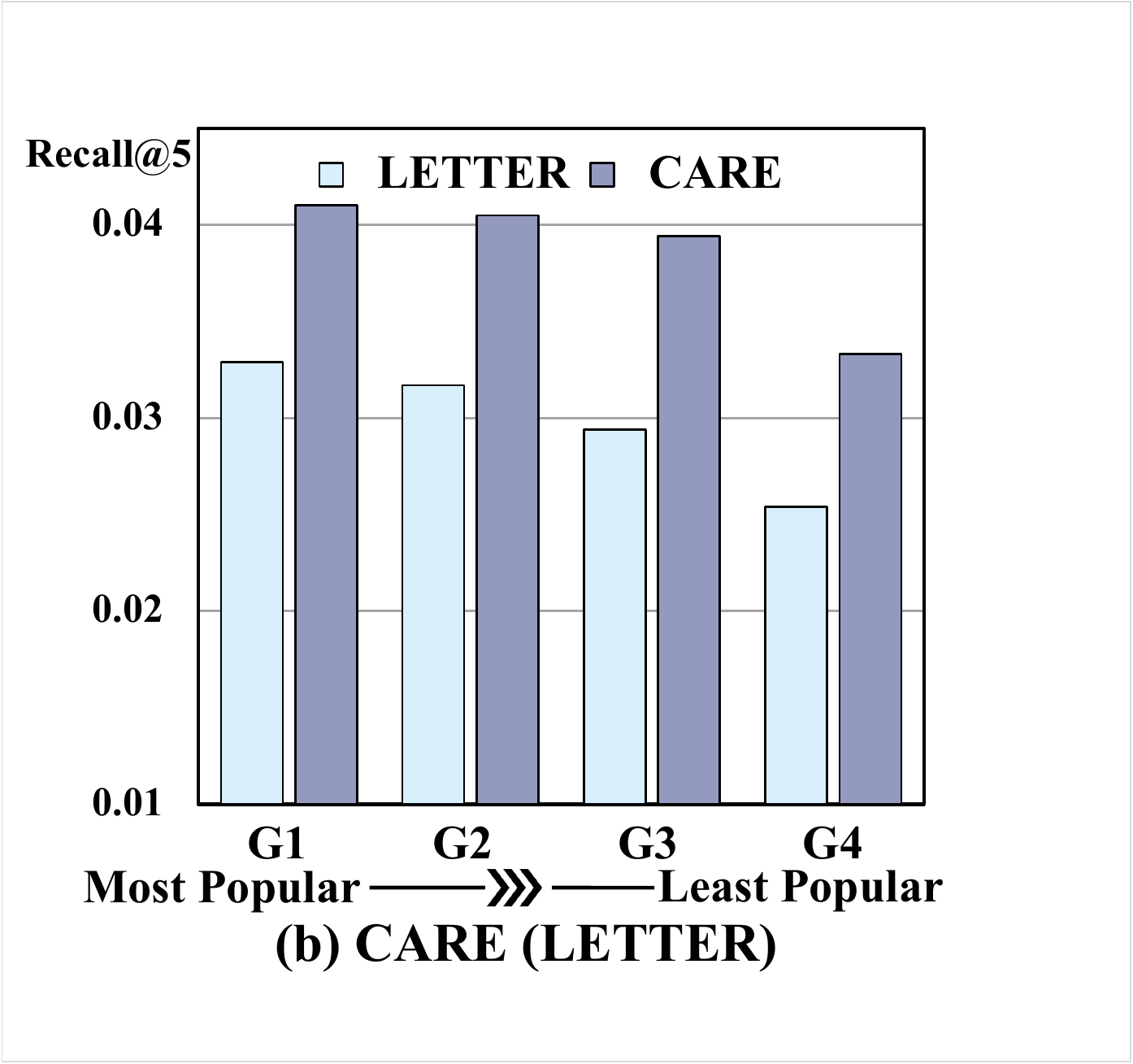}} 
\caption{Performance on user groups on Games dataset. The groups are sorted by interaction numbers. }
  \label{fig:user_group}
\end{figure}

\subsubsection{\textbf{Ablations Study on MicroLens}}\label{app:ablation_study}

\begin{figure}[h]
\vspace{-0.2cm}
\setlength{\abovecaptionskip}{-0.15cm}
\setlength{\belowcaptionskip}{-0cm}
  \centering 
  \hspace{-0.105in}
  \subfigure{
    \includegraphics[width=0.48\linewidth]{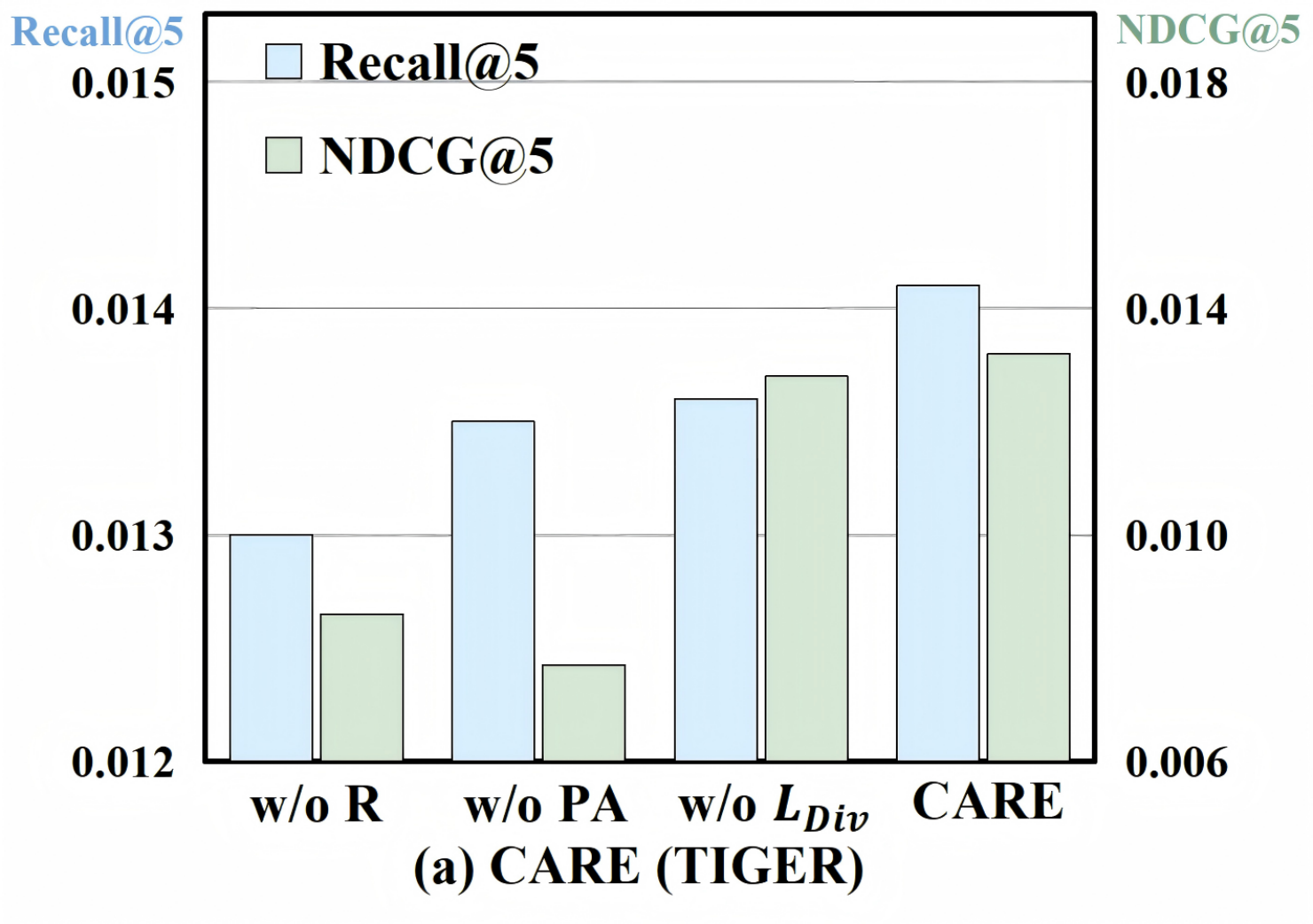}} 
  \hspace{-0.105in}
  \subfigure{
    \includegraphics[width=0.48\linewidth]{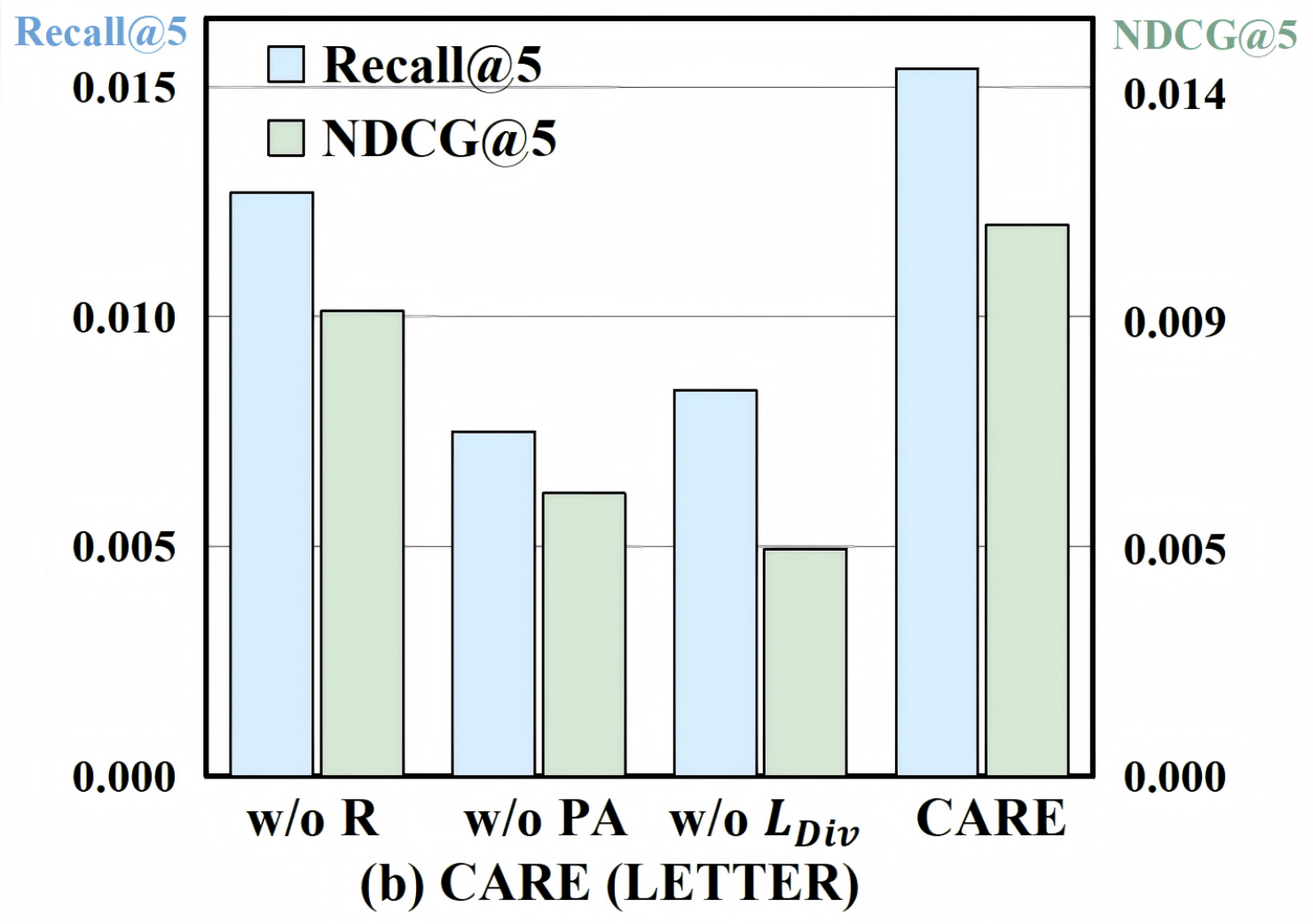}} 
\caption{Ablation study on MicroLens dataset. ``PA'' denotes ``Progressive Attention''. }
  \label{fig:Ablation_analysis2}
\end{figure}

The additional ablation study results on MicroLens are reported in Figure~\ref{fig:Ablation_analysis2}, where we have similar observations to those on Games dataset (\cf Section~\ref{sec:ablation}). 

\subsubsection{\textbf{Effect of Diversity Loss Strength $\bm{\alpha}$}}\label{app:effect_of_diversity} 
\begin{figure}[t]
\vspace{-0.2cm}
\setlength{\abovecaptionskip}{-0.15cm}
\setlength{\belowcaptionskip}{-0.3cm}
  \centering 
  \subfigure{
    \includegraphics[width=0.47\linewidth]{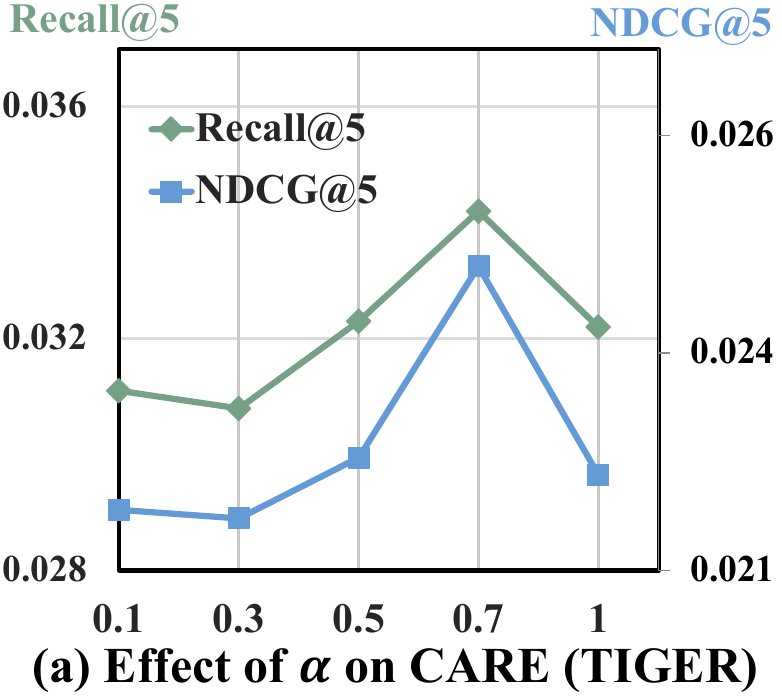}} 
  \subfigure{
    \includegraphics[width=0.47\linewidth]{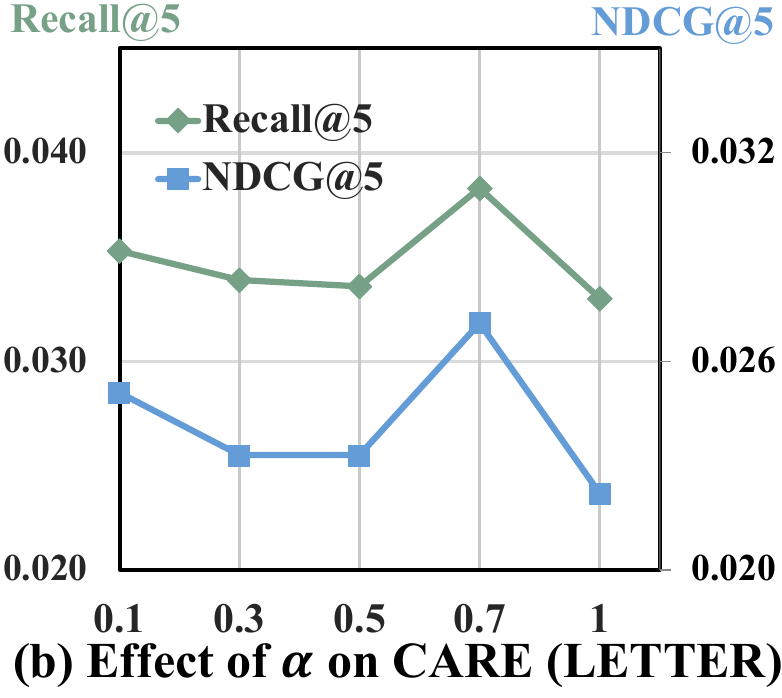}} 
\caption{Effect of $\bm{\alpha}$ on Games dataset. }
  \label{fig:DL_analysis}
\end{figure}

We vary $\alpha$ from 0.1 to 1 of CARE on the Games dataset as shown in Figure~\ref{fig:DL_analysis}. 
where the x-axis represents $\alpha$, y-axis represents Recall@5 and NDCG@5. 
From the results, we can find that gradually strengthening the penalty on similar queries tends to improve recommendation quality (from 0.1 to 0.7). However, too much emphasis on diversity might inversely hurt the optimization of recommendation capability, thus degrading the performance ($\alpha=1$). Empirically, we find that setting $\alpha=0.7$ yields the best performance. 

\subsubsection{\textbf{Additional Results on Parallel GR}}\label{app:parallel_GR}

The additional results of parallel GR on Toys and MicroLens datasets are presented in Figure~\ref{tab:overall_main2_toys_microlens}. In addition to the similar observations from that on Games and Sports (Table~\ref{tab:overall_main2}), we find that instantiating CARE on SETRec sometimes might yield inferior performance \wrt NDCG@$K$. One possible reason might be the characteristics of the grounding mechanism employed in SETRec. 
Specifically, the parallel generated item tokens are grounded to the item by linearly combining the scores from the CF token and the semantic tokens. 
Since the semantic tokens share the same weights in combination, the additional reasoning of CARE might lack diversity and limit the distinctions between each token's semantics, thereby hurting the relative ranking of the recommendations.

\subsubsection{\textbf{Further Analysis of Baseline Performance}}\label{app:overall_performance_analysis}
We present the detailed analysis of the baseline comparisons in Table~\ref{tab:overall_performance}: 
\begin{itemize}[leftmargin=*]
    \item Among the two generative methods, LETTER typically achieves better performance than TIGER,This is primarily attributed to two reasons: 1) During the learning of semantic IDs, LETTER incorporates collaborative filtering (CF) information, which encourages semantic IDs to simultaneously capture the rich semantics and co-occurrence patterns of items. Consequently, items with similar interactions acquire similar semantic IDs. 2) To foster more diverse item generation, LETTER promotes token utilization within semantic IDs, thereby enabling the model to enhance generation diversity by producing a wider variety of tokens.
    
\end{itemize}

\subsubsection{\textbf{Comparison of Efficiency Between CARE instantiated on TIGER and LETTER}}\label{app:efficiency_analysis}
An interesting finding from Table~\ref{tab:Efficiency_Analysis} is that LETTER requires less inference time than TIGER. 
We hypothesize that the reason for this is the more concentrated token distribution in TIGER's semantic IDs, whereas the semantic IDs in LETTER exhibit greater diversity~\cite{wang2024learnable}. This discrepancy leads to a less balanced Trie structure for semantic ID generation in TIGER. Consequently, the frequent inability to find a valid next token triggers numerous complex backtracking operations during beam search, which leads to significant CPU-side overhead and substantially increases the inference time.

\subsection{Detailed Experimental Settings}

\subsubsection{\textbf{Detailed Description for Baselines}}
We present the detailed descriptions of the traditional baselines below. 
1) \textbf{SASRec}~\cite{kang2018self} is one of the most representative sequential recommendation method, which 
adopts a unidirectional transformer model to capture sequential patterns in user behavior. 
2) \textbf{GRU4Rec}~\cite{hidasi2016session} utilizes recurrent networks for sequential recommendation. 
3) \textbf{CASER}~\cite{tang2018personalized} extracts sequential features from item ID inputs via convolutional filters.

\subsubsection{\textbf{Implementation Details of Baselines}}\label{app:implementation_detail}
For traditional baseline models(\textbf{SASRec}~\cite{kang2018self}, \textbf{GRU4Rec}~\cite{hidasi2016session}, \textbf{CASER}~\cite{tang2018personalized}), we search the learning rate in the range of $\{1e^{-3},1e^{-4}\}$ and the decay in the range of $\{1e^{-4},1e^{-5},1e^{-6}\}$. 

For the other baseline models, all learning rates are constrained within the range of $\{1e^{-3},5e^{-4},3e^{-4},1e^{-4}\}$. In addition, we also tune the hyperparameters of these baselines according to the recommendations in the original papers.
1) \textbf{RPG}~\cite{hou2025generating}: we use \textbf{sentence-t5-base}~\cite{ni2022sentence} as the semantic encoder. We tune the temperature $\tau \in \{0.07\}$ and semantic ID length $m \in \{32\}$.
2) \textbf{HSTU}~\cite{zhai2024actions}: To ensure the fairness of the experiments, we increase the parameter count of HSTU to 0.5B by enlarging the encoder QKV dimension to 896 and increasing the number of attention heads to 14.
3) \textbf{SETRec}~\cite{lin2025order}: We select $n_sem$ and $\beta$ from $\{1,2,3,4\}$ and 
$\{0, 0.1, 0.2, 0.3, 0.4, 0.5, 0.6, 0.7, 0.8, 0.9, 1.0\}$, 
where $n_sem$ denotes the number of semantic queries and $\beta$ denotes the hyperparameter of the fusion weight between collaborative filtering signals and semantic signals.
4) \textbf{ReaRec}~\cite{tang2025think}: Inspired by ReaRec, we implement ReaRec using SETRec. Specifically, we set $n_{sem}$ in SETRec to $0$ to retain only the collaborative filtering signals, while keeping the other hyperparameters identical to those of SETRec.
5) \textbf{SPRec}~\cite{gao2025sprec}: the total number of iterations is set to 5, with each SFT and DPO phase trained for 1 epoch.
\begin{table}[h]
\setlength{\abovecaptionskip}{0cm}
\setlength{\belowcaptionskip}{0cm}
\caption{Datasets statistics. ``Int.'' denotes ``interactions''. }
\setlength{\tabcolsep}{3mm}{
\resizebox{0.48\textwidth}{!}{
\begin{tabular}{lcccc}
\hline
 & \multicolumn{1}{l}{\textbf{\# Train Int.}} & \multicolumn{1}{l}{\textbf{\# Valid Int.}} & \multicolumn{1}{l}{\textbf{\# Test Int.}} & \multicolumn{1}{l}{\textbf{\# Item}} \\ \hline
\textbf{Games} & 201,612 & 25,201 & 25,202 & 11,036 \\
\textbf{Sports} & 181,476 & 22,684 & 22,685 & 16,002 \\
\textbf{Toys} & 112,754 & 14,094 & 14,095 & 11,251 \\
\textbf{MicroLens} & 60,100 & 9,035 & 9,120 & 5,326 \\ \hline
\end{tabular}
}}
\label{tab:dataset_statistics}
\end{table}

To implement the autoregressive GR backbone (TIGER and LETTER), we follow the settings in~\cite{wang2024learnable}. 
Specifically, for item tokenization, we use RQ-VAE to present each item with a semantic ID of length $l=4$. For the generative recommender, we use Qwen2.5-0.5B as the backbone, and search the learning rate in the range of $\{1e^{-3},5e^{-4},3e^{-4},1e^{-5}\}$, respectively.


{
\tiny
\bibliographystyle{ACM-Reference-Format}
\balance
\bibliography{bibfile}
}

\end{document}